\documentclass[letterpaper, twocolumn, 10pt]{article}
\usepackage{usenix-2020-09, outlines}

\usepackage{main_macros}

\newif\ifsavespace
\savespacefalse

\ifsavespace
\usepackage{etoolbox}
\makeatletter
\patchcmd{\maketitle}
	{\@maketitle}
	{\@maketitle\vspace{-5em}} 
	{}
	{}
\makeatother
\fi

\begin{document}
\date{}

\title{\Large \bf Application-Defined Receive Side Dispatching on the NIC}

\ifarxiv
\author{
{\rm Tao Wang$^\dagger$}
\and
{\rm Jinkun Lin$^\dagger$}
\and 
{\rm Gianni Antichi$^{\ddagger}$}
\and
{\rm Aurojit Panda$^\dagger$}
\and
{\rm Anirudh Sivaraman$^\dagger$}
\and 
{$^\dagger$New York University~~~$^\ddagger$Politecnico di Milano and Queen Mary University of London}
}
\else
\author{Draft, \ref{TotPages} pages total}
\fi

\maketitle

\begin{abstract}
Application layer (L7) processing is increasingly implemented in proxies (e.g., Envoy) to simplify administration and management.  
However, prior work has observed that this reduces application performance and increases resource requirements.
The reason is that moving logic out of the application required duplicating some computation and additional inter-process communication.
This paper describes \sysname, a system that moves L7 dispatch (a function implemented by all L7 proxies and affects all messages received by an application) to a NIC that is on the application's communication path. 
Unfortunately, the data formats and protocols used by modern applications pose a challenge when moving L7 dispatch to NICs.
Consequently, when designing \sysname we had to rethink not just the NIC hardware, but also how applications encode data sent over the network.
We prototyped \sysname using a 100GbE FPGA NIC, and show that for real-world applications \sysname can achieve 6.6$\times$ to 7.15$\times$ higher throughput compared to software proxies. 

\end{abstract}

\section{Introduction}
\label{sec:introduction}

Datacenter operators are increasingly moving application layer (L7) processing (\eg service discovery, L7 load balancing~\cite{nginx_http_lb}, etc.) from the application itself into software proxies (\eg Envoy~\cite{envoy_proxy}, Nginx~\cite{nginx_http_lb}, ServiceRouter~\cite{service_router}, etc.), as it simplifies application management and deployment~\cite{mrpc}. 

In particular, L7 dispatch is a very important proxy function that impacts the performance of nearly all messages. This is because it must be run by the proxy on every message to determine what thread (or process) should handle the message itself, and requires that the proxy analyze message contents. Popular implementations have high overheads for this function, for example, using Istio~\cite{istio_service_mesh} increases latency by 269\%~\cite{meshinsight} and CPU usage by 163\%. Consequently, many prior efforts~\cite{envoy_proxy, istio_service_mesh, grpc_over_pb, linkerd, nginx_http_lb, service_router, airbnb_synapse, aws_microservice, yoda, uber_service_mesh, netflix_ribbon} have focused on improving its performance. 

In this paper, we ask \textit{can we delegate L7 dispatch to the NIC hardware which is located on the application's communication path so as to improve performance?} While prior work~\cite{protoacc, cereal, cerebros, optimus_prime} has offloaded other L7 processing to NICs, as we discuss later in \S\ref{sec:related_work} none of these approaches suffice for offloading L7 dispatch.

\begin{figure}
    \centering
    \includegraphics[width=0.8\linewidth]{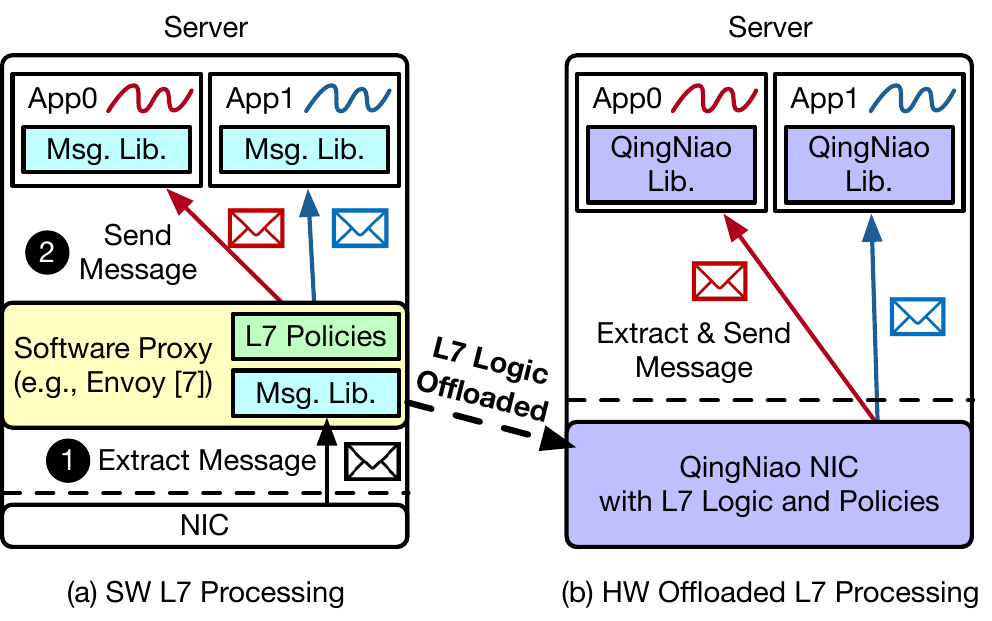}
    \caption{Architectural comparison of L7 processing implemented in software and hardware (\ie \sysname).
    \label{fig:arch-overview}}
\end{figure}

To understand how to do this, we must first understand what in the current architecture (Figure~\ref{fig:arch-overview}a) leads to these overheads. Software L7 dispatch requires performing two steps: (1) extracting application messages from network packets and applying policies (\eg authentication) on the extracted data; (2) using inter-process communication (IPC) to send the application message to the thread that processes it. Both steps add overheads (\S\ref{sec:background}), and thus our approach seeks to alleviate the impact of both.

Our design, \sysname (Figure~\ref{fig:arch-overview}b),
requires applications to adopt a new encoding scheme (\ie serialization format) for data sent over the network, and implements L7 dispatch logic directly on the NIC, thus avoiding expensive L7 processing alongside IPC to be performed on the CPU host.

To realize this vision, we had to address two main challenges:

The first, was a mismatch between the size of application messages and network packet sizes: an application message can be several megabytes in size, exceeding the MTU of most networks. Consequently, a single application message is often \emph{segmented}, which is split across multiple packets. Software-based proxies rely on the OS network stack to reassemble packet payloads into an application message before processing. However, implementing segmentation reassembly on hardware requires complicated logic and large amounts of memory~\cite{auto_nic_offload}. Therefore, we designed our encoding scheme (\S\ref{ssec:design_qn_sw}) to avoid requiring reassembly: each \sysname packet contains a \emph{message ID}, and the first packet of a message contains all of the fields that can be used to dispatch the message. This design allows the NIC to process individual packets without reassembly: the NIC determines and caches a dispatch decision when it receives the message's first packet, and all subsequent message packets are dispatched using the cache.

The second, was the presence of variable-length fields (\eg strings or vectors) within the application messages. Existing match-action pipelines are designed to process fixed-length fields (\eg IP addresses), and this affects both how match rules are specified (using exact values or bit masks) and processed. We needed to change both in our setting: we adopt a skip-and-match based rule specification (\S\ref{ssec:qn_hw_arch}) to specify matching rules over variable length fields, and design a hardware match engine (\S\ref{sssec:qn_rsd}) that can efficiently match packet contents using these rules.

We have prototyped \sysname on an Xilinx Alveo U250 FPGA board~\cite{au250} and integrated with a 100GbE FPGA SmartNIC by extending Corundum~\cite{corundum}.\footnote{Our prototype is open source at \arxivrepo.} Our prototype consumes about 12.88\% LUT, 8.22\% LUTRAM, and 7.31\% BRAM of the total resources. It achieves 6.6$\times$ to 7.15$\times$ higher throughput and 72.46\% to 74.62\% lower latency compared to a state-of-the-art software L7 dispatch based on eBPF~\cite{spright}. 

\section{Overheads of L7 Dispatch in Software}
\label{sec:background}

\begin{table}[!t]
\begin{center}
\begin{tabular}{lr}
\toprule
\textbf{Component} & \textbf{Time} \\
\midrule
\textbf{HTTP Processing} & 27 \si{\micro\second} \\
\textbf{Interprocess Communications} & 15.23 \si{\micro\second} \\
\cmidrule(lr){2-2}
Total & 42.23 \si{\micro\second} \\
\bottomrule
\end{tabular}

    \caption{Dissection of additional latency brought by Envoy. 
    \label{tab:latency_dissection}}
\end{center}
\end{table}

We start by measuring the overheads on response latency for performing L7 dispatch in software. 
To do so we compare response latency for FastHTTP~\cite{fasthttp} when running standalone to a deployment where Envoy~\cite{envoy_proxy}\footnote{We used version v1.21.0.}, a widely adopted L7 proxy, is used to implement L7 dispatch. 
We use the same FastHTTP configuration in both cases. For the Envoy-based deployment, we use one Envoy instance and two FastHTTP instances, and pin Envoy and FastHTTP to different cores. In both cases, we ensure there is no request queuing and report average latency across 100\si{K} iterations. The experimental setup is detailed in and consistent with \S\ref{sec:evaluation}.

In our setting, FastHTTP when run without Envoy has a response latency of $45$\si{\micro\second}. Running it with Envoy increases this latency to $87.23$\si{\micro\second}, an increase of 91.9\%. We used BCC's funclatency~\cite{bcc_funclatency} to understand the source of these overheads (Table~\ref{tab:latency_dissection}). We found that Envoy's HTTP parsing contributed approximately 27\si{\micro\second}, while IPC between Envoy and FastHTTP added approximately 15\si{\micro\second}. To put this breakdown in context, HTTP parsing within Envoy takes 60\% of the time FastHTTP takes to process a request, while IPC takes 33.9\% of the time FastHTTP takes for processing requests. Thus both contribute to the additional response latency, and need to be addressed.

Not only does Envoy increase response latency, it also decreases throughput. We measured this using wrk~\cite{wrk}, and found that adding Envoy reduces throughput by over $9\times$: FastHTTP without Envoy can process 123.65 \si{K}rps, while with Envoy it can only process 13.13 \si{K}rps.

These results show the importance of eliminating software L7 dispatch overheads by offloading it to hardware, and we describe our approach for doing so next.
\section{\sysname}
\label{sec:design}

\begin{figure}[!t]
    \centering
    \includegraphics[width=0.7\linewidth]{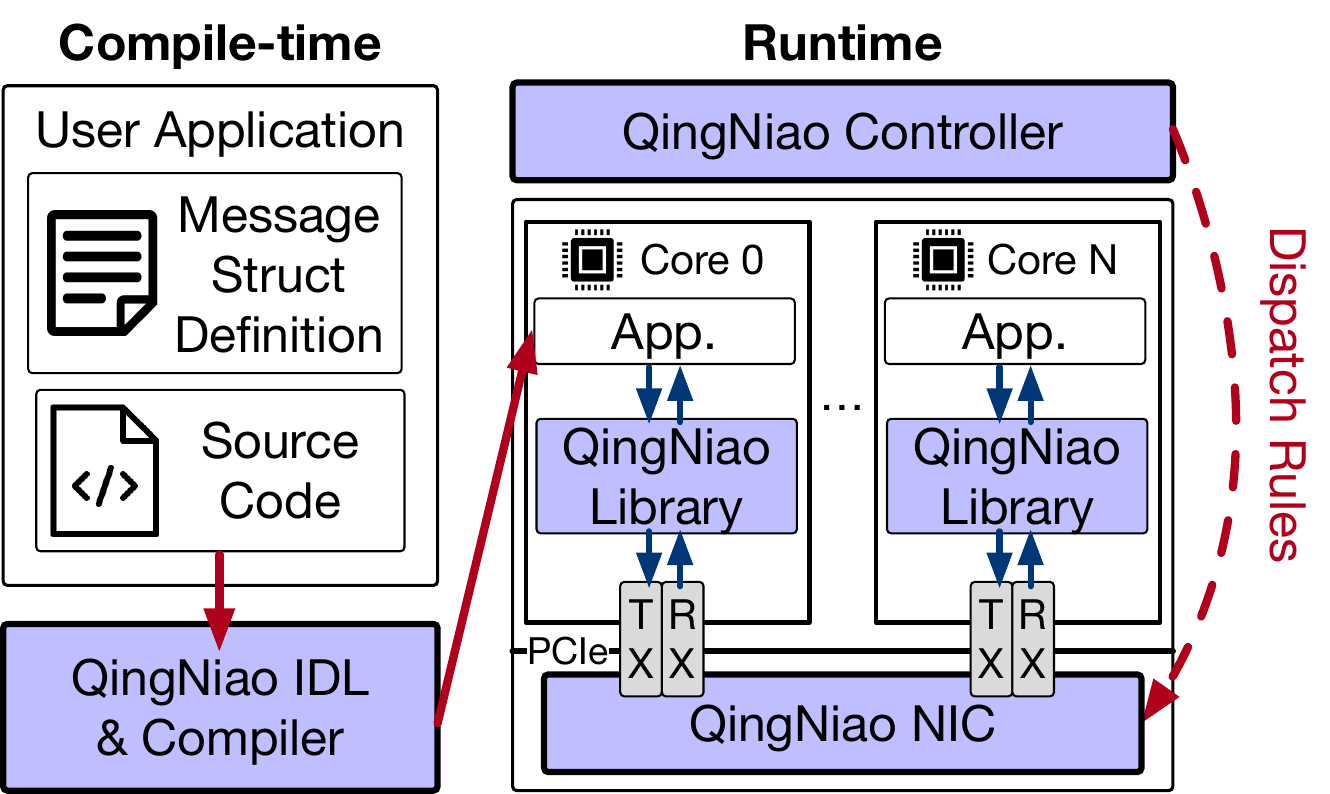}
    \caption{\sysname's overview. \sysname's design consists of the colored components.
    \label{fig:qn_overview}}
\end{figure}

\sysname's goal is to offload L7 dispatch to a NIC, and thus eliminate the overheads of software dispatchers. We had two primary goals when designing \sysname: (1) \textit{generality}, allowing \sysname to be used by different L7 message definitions and with different L7 dispatch rules; (2) \textit{hardware resource efficiency}, allowing \sysname to be implemented on a wide variety of NICs and combined with other types of offloads.

\sysname (Figure~\ref{fig:qn_overview}) consists of the three components: 
(1) \textbf{\sysname software (\S\ref{ssec:design_qn_sw})}, which provides interfaces for the applications: a \sysname Interface Definition Language (IDL) and its compiler to customize L7 message structs for achieving \textit{generality}, a runtime API library to interact with the underlying \sysname NIC; 
(2) \textbf{\sysname hardware (\S\ref{ssec:qn_hw_arch})}, which achieves \textit{resource efficiency} by buffering no packets, thus avoiding excessive memory footprint and complex on-NIC buffer management for L7 message reassembly;
(3) \textbf{\sysname controller}, which manages on-NIC \sysname L7 dispatch rules (\S\ref{sssec:qn_rsd}) at runtime.

Data layout is the core technique that enables \sysname. When segmenting an application message that spans multiple packets, \sysname lays out data so that: (1) portions relevant to L7 dispatch appear in the message's first packet, (2) and each packet carries this message's unique ID.
As a result, \sysname allows to dispatch every packet: a dispatch decision can be made on the message's first packet, which can be looked up for subsequent packets using the same message ID.

To make prototyping feasible, our current design does not use TCP as a transport protocol, and instead uses a simpler \sysname protocol (QNP) described in \S\ref{ssec:impl_qn_sw}. As we discuss later in \S\ref{sec:discussion}, our design choices can be used with TCP or other transport protocols.

\subsection{\sysname in Action}
\label{ssec:qn_prog_model_example}

\begin{figure}[!t]
    \centering
    \begin{subfigure}[t]{0.47\linewidth}
        \includegraphics[width=\textwidth]{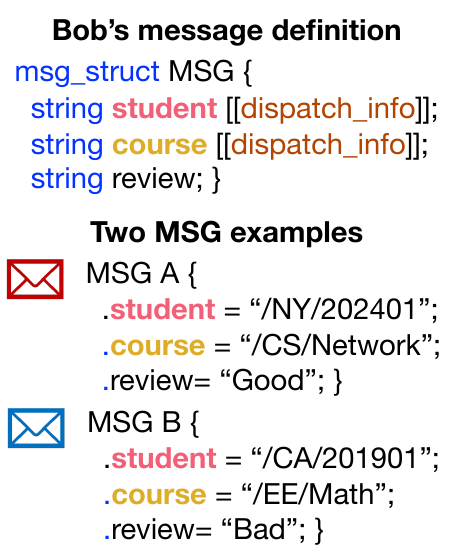}
        \caption{Example of messages.
        \label{fig:bob_req_def}}
    \end{subfigure} \hfill
    \begin{subfigure}[t]{0.47\linewidth}
        \includegraphics[width=\textwidth]{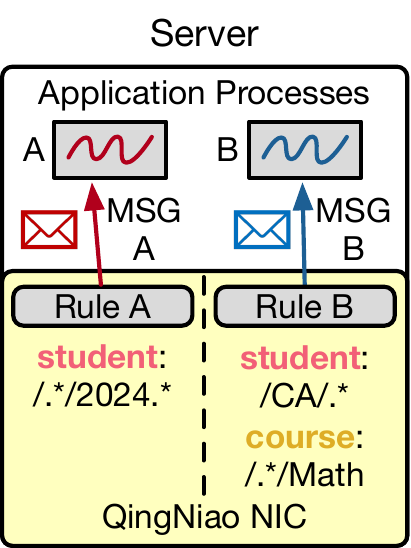}
        \caption{Example of dispatch rules.
        \label{fig:bob_dispatch_rules}}
    \end{subfigure} \hfill
    \caption{Example of Bob's message struct definition and dispatch rules.
    \label{fig:qn_bob_example}}
\end{figure}

\begin{figure}[!t]
    \centering
    \includegraphics[width=0.8\linewidth]{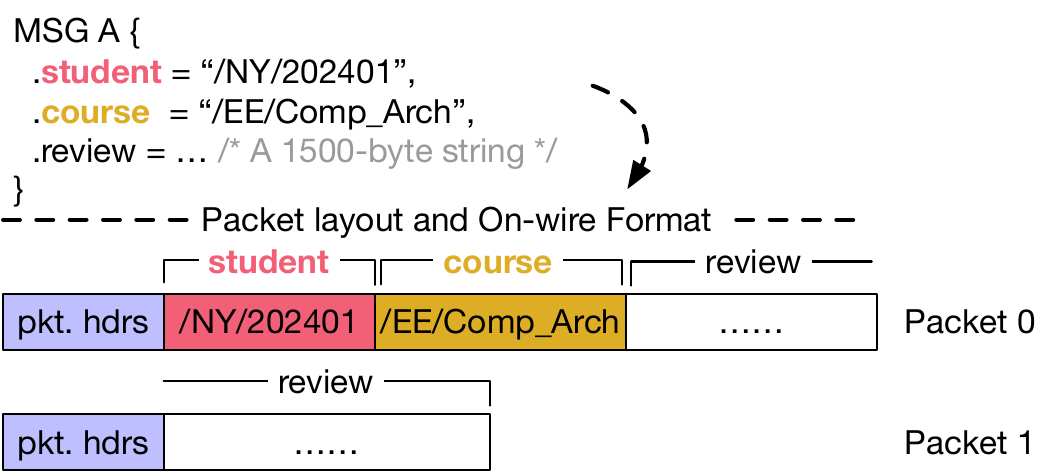}
    \caption{An example of packet layout and on-wire encoding of one Bob's message.
    \label{fig:bob_pktlayout}}
\end{figure}

Before diving into the details of \sysname, we first present an example of how to build and deploy a \sysname application. We use this example throughout the rest of this section to illustrate \sysname's design.

Suppose Bob, a developer at a university, wants to build and deploy a course review application. The application needs to be able to track reviews for courses offered by all of the university's departments.
Bob starts by defining his application's interface. The application processes messages (Figure~\ref{fig:bob_req_def}) that consist of three variable-length string fields:
(1) \textit{``student''} with the format ``/<campus>/<id>''; (2) \textit{``course''} with the format ``/<department>/<course>''; and (3) \textit{``review''}.

Bob wants to ensure that the application can scale to all users at the university, and span multiple cores. His design for scaling requires that he launches multiple processes that execute the application, and shard reviews (\eg by student ID or department) across these processes. Because data is sharded, he must ensure that messages are forwarded to the correct shard, and he must use an L7 dispatcher.

To use \sysname as an L7 dispatcher, Bob first defines the application messages (Figure~\ref{fig:bob_req_def}) using the \sysname IDL (\S\ref{ssec:design_qn_sw}). When defining application messages, Bob uses the \texttt{dispatch\_info} annotation to indicate the set of fields that can be used for L7 dispatch. Then, Bob writes the course review application using the \sysname library (\S\ref{ssec:design_qn_sw}). Next, Bob must decide what L7 dispatch rules he should use. We show two example dispatch rules in Figure~\ref{fig:bob_dispatch_rules}: Rule A directs any messages where the \textit{``student''} field matches the pattern \texttt{/*/2024} to process A, while Rule B directs messages where the \textit{``student''} matches \texttt{/CA/*} and \textit{``course''} field matches \texttt{/*/Math} to process B.

After creating the application and determining dispatch rules, Bob deploys all of the application processes on a server equipped with a \sysname NIC. Then Bob uses \sysname's controller (Figure~\ref{fig:qn_overview}), to configure dispatch rules. Once this is done, \sysname forwards messages to the configured processes, thus performing L7 dispatch.

\subsection{\sysname Software Stack}
\label{ssec:design_qn_sw}

We discuss the design of \sysname software and defer the detailed implementation to \S\ref{ssec:impl_qn_sw}.

\Para{\sysname IDL and compiler.} Similar to protobuf~\cite{protobuf}, \sysname provides an IDL and compiler for customizing message structs. Additionally, it provides a \textit{dispatch\_info} attribute specifier to mark fields used for L7 dispatch as in Figure~\ref{fig:bob_req_def}.

\Para{Application Programming Interfaces (APIs).} 
\sysname presents a minimalistic programming model to applications, namely \textit{send\_msg()} and \textit{recv\_msg()} APIs in \sysname library. 

To send messages to the underlying network, applications use the \textit{send\_msg()} API, which lays out and serializes message fields.
Specifically, for the layout, \sysname library arranges the message fields such that the fields marked with ``\textit{dispatch\_info}'' appear before the other fields. Then the message is segmented into multiple packets if necessary, and the ``\textit{dispatch\_info}'' fields are guaranteed to appear complete in the first packet. In the packet header, each packet will carry the message's unique ID and its in-message packet sequence number. 
As we will show in \S\ref{ssec:qn_hw_arch}, \sysname hardware offloads L7 dispatch by leveraging this packet layout to avoid on-NIC packet buffering and message reassembly.

For the serialization, \sysname library uses the serialization functions---generated by IDL compiler---to translate message fields into a Type-Length-Value (TLV~\cite{tlv_encoding}) format, which is effective in expressing variable-length L7 values and widely used~\cite{protobuf}. 
Take Figure~\ref{fig:bob_pktlayout} as an example, Bob's message is laid out and serialized into two packets.

On a server equipped with \sysname NIC, each application process is allocated with a set of TX/RX queues (Figure~\ref{fig:qn_nic_arch}, \S\ref{ssec:qn_hw_arch}). To receive messages from the underlying network, applications use the \textit{recv\_msg()} API. 
The \sysname NIC and library will ensure the packets with the same message are all received in order and deserialized into the original message, which is delivered to the application.

\subsection{\sysname Hardware Architecture}
\label{ssec:qn_hw_arch}

\begin{figure}[!t]
    \centering
    \includegraphics[width=1.0\linewidth]{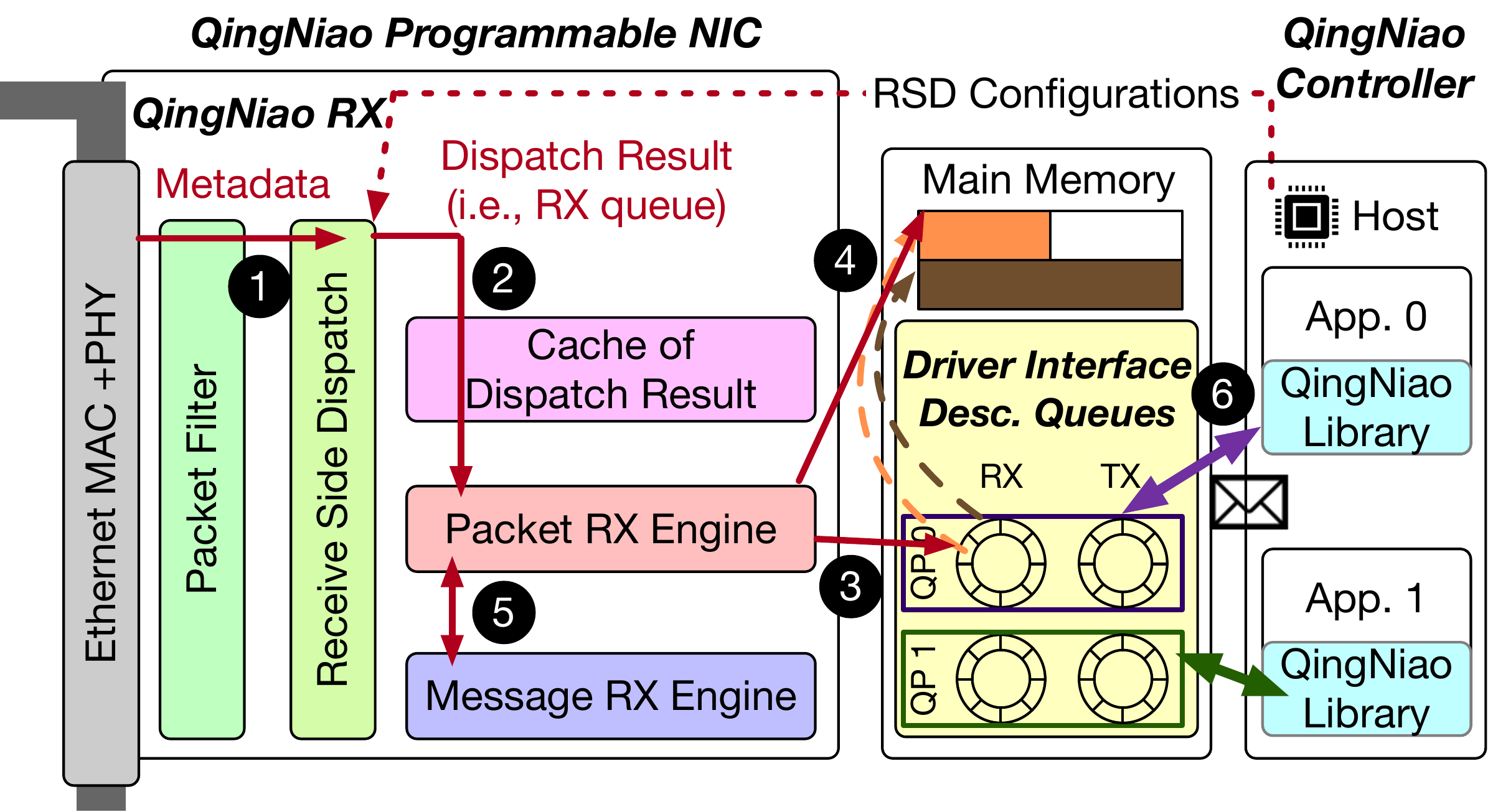}
    \caption{\sysname's RX data path overview. \label{fig:qn_nic_arch}}
\end{figure}

Since \sysname primarily works on the RX path to dispatch messages, we first describe how the RX path works in \sysname NIC. The TX path is simpler, merely requiring the NIC to send out packets produced by the \sysname library.
Then, we detail the design of the Receive Side Dispatch (RSD, \S\ref{sssec:qn_rsd}) hardware module, which is designed to carry out L7 dispatch in \sysname.

As in~\S\ref{ssec:design_qn_sw}, for a multi-packet message, \sysname library ensures that (1) \textit{dispatch\_info} fields are placed in the first packet; (2) each packet carries this message's ID.

Based on this packet layout, \sysname NIC is designed to match \textit{dispatch\_info} fields against configured rules and cache the dispatch result for subsequent packets of the same message ID.

By leveraging this, \sysname NIC avoids buffering packets and reassembling messages on NIC, which is shown to be resource-consuming~\cite{ttpoe, hotchip34_ttpoe}. Instead, to achieve the design goal of hardware resource efficiency, \sysname NIC only incurs minimal on-NIC memory footprint: it only maintains per-message states (\ie descriptor for DMAing packet payloads, sequence numbers of received packets, timer for message expiration, cached dispatch result) for each active message. 
To this end, \sysname uses the following mechanisms:

\Para{Separation of packet and message delivery.} 
\sysname does not buffer packets: 
each received packet is immediately DMAed to the host memory and the host will be notified later when all packets of a message are received.
For ease of exposition, here we assume in-order packet arrival within a message. We discuss how \sysname handles out-of-order packets later. 

As shown in Figure~\ref{fig:qn_nic_arch}, on the {\it RX} path, 
\begin{CompactItemize}
\item \circled{1} a packet filter parses out packet metadata (\ie message ID, length, and packet sequence number) and
inputs it to RSD (\S\ref{sssec:qn_rsd}) along with the packet to compute the \circled{2} dispatch result (\ie RX queue) for the message.
\item The result is cached for this message ID and used by Packet Receive Engine to \circled{3} fetch a descriptor from the target RX queue, then \circled{4} packet content is DMAed to the host, by using the fetched descriptor and the packet's in-message sequence number to determine its offset in the descriptor-pointed memory region. \sysname uses per-message descriptor and ensures the pointed memory region is large enough for the message itself.
\item \circled{5} Message Receive Engine tracks the following states per message: the fetched descriptor, an indicator of received and missing packets to show whether a message is fully received, and a timer when the last packet of the message was received. These states are essential for \sysname's correctness and resource efficiency: the descriptor is kept for the subsequent packets within the same message, and the per-message timer is used to notify \sysname to reclaim the corresponding on-NIC states and at-host memory once it expires.
\item \circled{6} When all packets of the message are received, \sysname NIC notifies the host of the successful message delivery. By calling \sysname \textit{recv\_msg()} API, applications can directly receive the deserialized messages from the allocated RX queues.
\end{CompactItemize}

\Para{Handling out-of-order packet arrival.}
As discussed in \S\ref{ssec:design_qn_sw}, \sysname guarantees that only the first packet within a message carries information for L7 dispatch. Then the dispatch decision will be cached for the subsequent packets within the same message, which means that the following packets in a message can not be dispatched until the first one comes. 
However, out-of-order packet arrivals may happen due to packet loss or network failures~\cite{vl2}. To avoid buffering every out-of-order packet on NIC, \sysname NIC simply discards packets from a message if its first packet was not seen before, and leaves it to the upper transport to deal with retransmissions. 
Once a message's first packet is seen and processed, out-of-order arrivals of subsequent packets can be simply addressed by accessing the dispatch result cached on NIC.
Such simplified design choices are usually adopted in the NIC design, including RDMA's go-back-N retransmission strategy~\cite{rdma_at_scale} and Tesla's TTPoE~\cite{ttpoe}, to avoid directly buffering packet contents on NIC.

\subsubsection{\sysname Receive Side Dispatch}
\label{sssec:qn_rsd}

\sysname Receive Side Dispatch (RSD) is the hardware module responsible for mapping messages to desired RX queues based on the specified L7 dispatch rules. We first describe the type of dispatch rules supported by \sysname, and then elaborate how RSD is designed to match incoming messages against configured dispatch rules.

\Para{\sysname dispatch rules based on skip-and-match.} As discussed in \S\ref{sec:introduction}, L7 dispatch usually requires matching message fields of byte strings against configured rules. To support this, one straightforward way is to implement on-NIC regex as it is effectively expressive for string matching~\cite{regex_wiki}.

Unfortunately, implementing reconfigurable regex matching based on Deterministic Finite Automatas (DFAs) using lookup tables~\cite{rmt_parser_design} is complex and challenging~\cite{hare} since the number of states of DFAs grows exponentially as the complexity of regex increases~\cite{dfa_state, intro_to_automata}.

To this end, \sysname instead adopts a simpler string matcher based on a \textit{skip-and-match} abstraction which is a relaxed form of regex. The rationale behind this is that dispatch rules often match on only portions of a string, \eg the first few bytes of an argument or an URL pattern.
Specifically, one skip-and-match consists of an alternating sequence of bytes that should be skipped and strings that should be matched. Specifically, skip-and-match provides two primitives \textit{Skip} and \textit{SkipUnitl}, where the numbers of skipped bytes are \textit{fixed} or \textit{variable}, respectively.

Recall Bob's dispatch Rule A (\ie \texttt{/.*/2024}) in Figure~\ref{fig:bob_dispatch_rules} as an example, it can be achieved by two skip-and-matches: (1) \textit{Skip} 0 bytes and Match \texttt{/}; (2) \textit{SkipUntil} a specified character (\eg `\texttt{/}') and Match \texttt{2024}. This rule matches any string that has the prefix starting with \texttt{/}, and ending with \texttt{/2024}. Concretely, it will match the string \texttt{/CA/20240919}, but not the string \texttt{/PA/202345}.

In practice, skip-and-match is sufficiently expressive to encode L7 dispatch rules given a known string format. For example, in an HTTP URL~\cite{http_url} string, a path component typically consists of a sequence of path segments separated by a slash (\eg \url{https://harrypotter.fandom.com/wiki/Hedwig}). Using skip-and-match's \textit{Skip} and \textit{SkipUntil} primitives can easily match desired portions of such strings as elaborated before.

\begin{figure}[!t]
    \centering
    \includegraphics[width=0.85\linewidth]{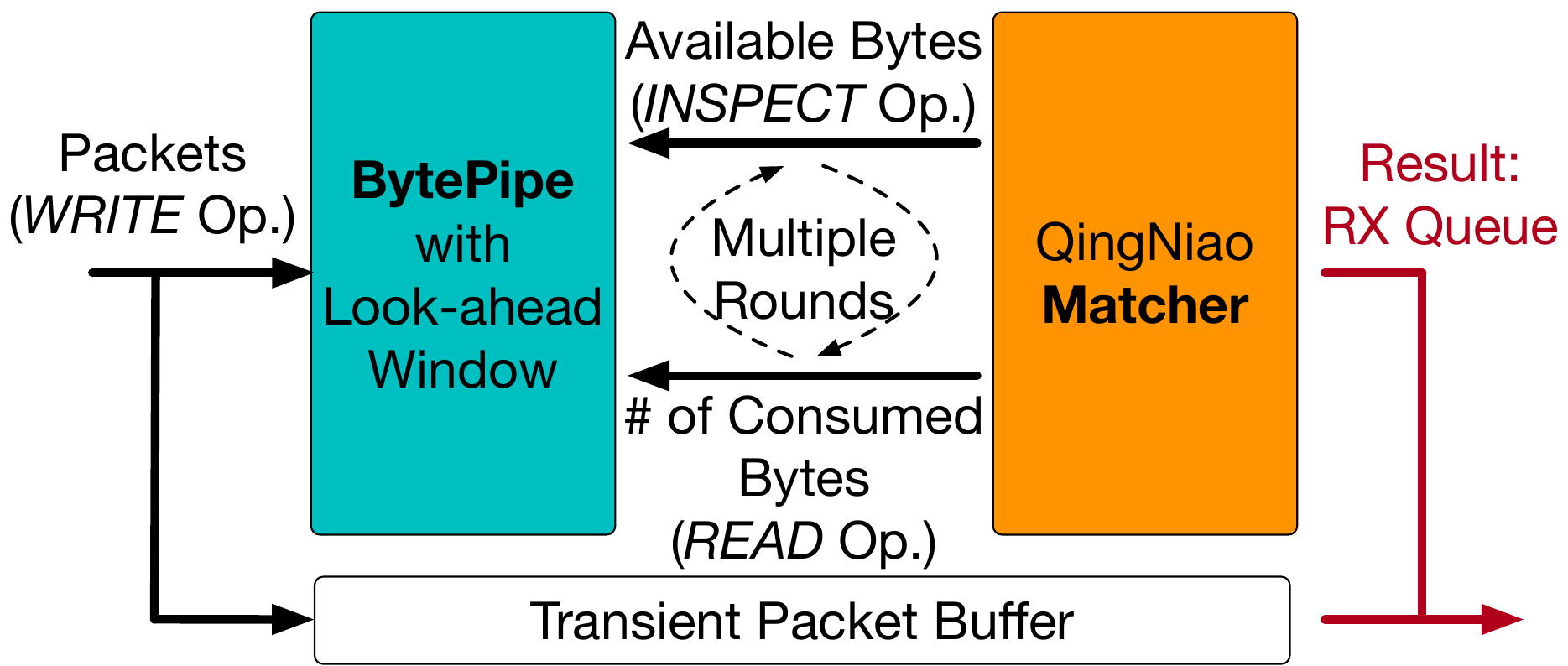}
    \caption{Receive Side Dispatch's components: BytePipe with look-ahead window and programmable Matcher. 
    \label{fig:rsd_overview}}
\end{figure}

\Para{How RSD works.} \sysname RSD extracts L7 message struct fields from packet contents and matches them against the configured dispatch rules to emit the dispatch result (\ie RX queue). To fulfill this, RSD should be able to (1) decode TLV-encoded message fields, which requires RSD to parse field type, length, and value in order; (2) move along the field's byte sequence and match it against the dispatch rules, consisting of skip-and-matches. They both require processing bytes serially. However, memory elements typically do not natively support serial access at byte granularity~\cite{xilinx_memory}.

Therefore, the RSD (Figure~\ref{fig:rsd_overview}) is designed with 2 parts: (1) a BytePipe, and (2) a programmable Matcher. BytePipe takes in the packets and serves as a serial transient byte stream for the Matcher as it presents the Matcher with the available bytes. Matcher will consume bytes from BytePipe to parse L7 message fields in the TLV format, and match them against configured dispatch rules. The original packet contents are also stored in the transient packet buffer, which will emit the packets along with the dispatch result to the downstream module.

Specifically, as shown in Figure~\ref{fig:rsd_overview}, to support serial parsing for the Matcher, BytePipe provides 3 operations: {\it read} and {\it write}---with a number of bytes as an argument---specify how many bytes are read out and written in, respectively; {\it inspect} checks bytes of a look-ahead window.
By interacting with BytePipe, Matcher is designed to carry out skip-and-match: it uses {\it read} operation to {\it skip} bytes and then instructs {\it inspect} operation to get a window of bytes, which is used to {\it match} against the bytes configured in dispatch rules.

Additionally, by instructing the packet filter to discard the traffic from applications under configuration, Matcher supports disruption-free reconfiguration, as shown in \S\ref{ssec:eval_microbench}, to program dispatch rules at runtime. 
In \S\ref{ssec:impl_qn_hw}, we describe an FPGA implementation of RSD without incurring host CPUs.


\section{Implementation}
\label{sec:implementation}

In this section, we detail the prototype of \sysname hardware and software stack, which includes the \sysname IDL compiler and library, and ensures the packet layout---required by \sysname hardware---in a \sysname protocol (QNP).
At last, two optimization techniques are introduced to improve \sysname hardware's throughput.

\subsection{\sysname Hardware Prototype}
\label{ssec:impl_qn_hw}

We prototype \sysname hardware using FPGA and integrate it into Corundum~\cite{corundum}, an open-source 100GbE NIC architecture on FPGA, using \verilogLOC lines of Verilog. \sysname hardware uses a 512-bit AXI-S~\cite{xilinx_axis} data width and runs at 250 \si{MHz}. The code builds on the base Corundum commit~\cite{corundum_base_commit}.


\begin{figure}[!t]
    \centering
    \includegraphics[width=1.0\linewidth]{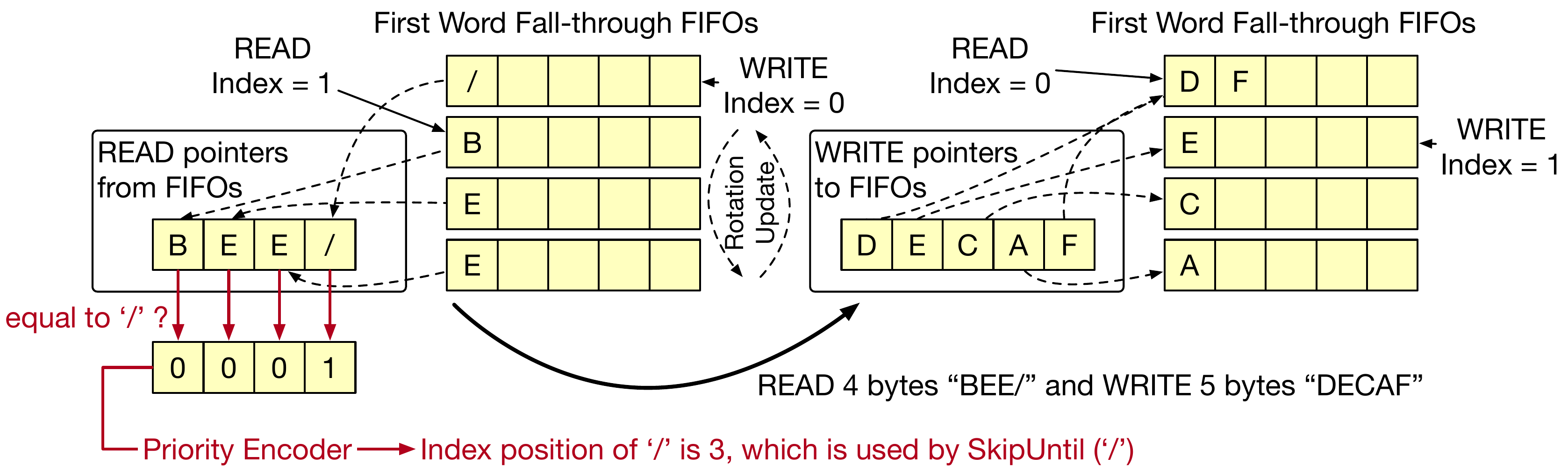}
    \caption{A BytePipe example with read and write index being 1 and 0 at the beginning, respectively. A {\it read} operation reads out 4 bytes (\ie ``BEE/'') and a following {\it write} operation writes in 5 bytes (\ie ``DECAF''). BytePipe leverages a priority encoder to indicate the position of the specified character used by SkipUntil.
    \label{fig:BytePipe}}
\end{figure}

\Para{Receive Side Dispatch.} As discussed in \S\ref{sssec:qn_rsd}, RSD mainly consists of \textit{BytePipe} and \textit{Matcher} modules, of which we show the implementations in the following.

\textit{BytePipe.} As in Figure~\ref{fig:BytePipe}, inspired by protoacc~\cite{protoacc}, BytePipe is implemented as a set of {\it parallel} First-Word Fall-Through FIFO queues (FWFT FIFOs), where the word size is 1 byte and the first byte at the head is immediately presented if it is not empty. This allows BytePipe to accept a simultaneous read/write of a sequence of bytes, which is capped by the number of parallel FWFT FIFOs. To support {\it read}, {\it write} and {\it inspect}, BytePipe keeps track of {\it read} and {\it write} index when executing {\it read} and {\it write} operations, respectively.

Take Figure~\ref{fig:BytePipe} as an example. For {\bf read}, BytePipe starts with the {\it read} indexed FIFO and iterates on parallel FIFOs to extract byte till the desired length. A 4-byte read operation starts from 1-indexed FIFO and BytePipe outputs \texttt{BEE/}. Note that reading from parallel FIFOs can be done simultaneously.
{\bf Inspect} works in the same way as {\it read}, except that it only exposes the bytes but does not consume them. At the same time, to support \textit{SkipUntil}, a priority encoder is used to indicate the index of the specified character (\eg, `/' in the example), which can be used as the number of bytes to be skipped.
For {\bf write}, BytePipe arranges the insertion of the byte to each FIFO starting from the {\it write} indexed FIFO. As shown in Figure~\ref{fig:BytePipe}, \texttt{DECAF} of 5 bytes will be spread out over all 4 FIFOs and two rounds of inserting bytes to FIFOs, where the 0-indexed FIFO will be written twice.

In our current prototype, the number of the parallel FWFT-FIFOs is set to 64, allowing it to support at most 64 bytes for each round, where each operation may take multiple rounds. BytePipe needs 3 cycles for the {\it read} and {\it write} round, while {\it inspect} operates immediately. The depth of each FIFO is set to 128, allowing it to store 8192 bytes at most transiently.

\begin{figure}[!t]
    \centering
    \includegraphics[width=0.8\linewidth]{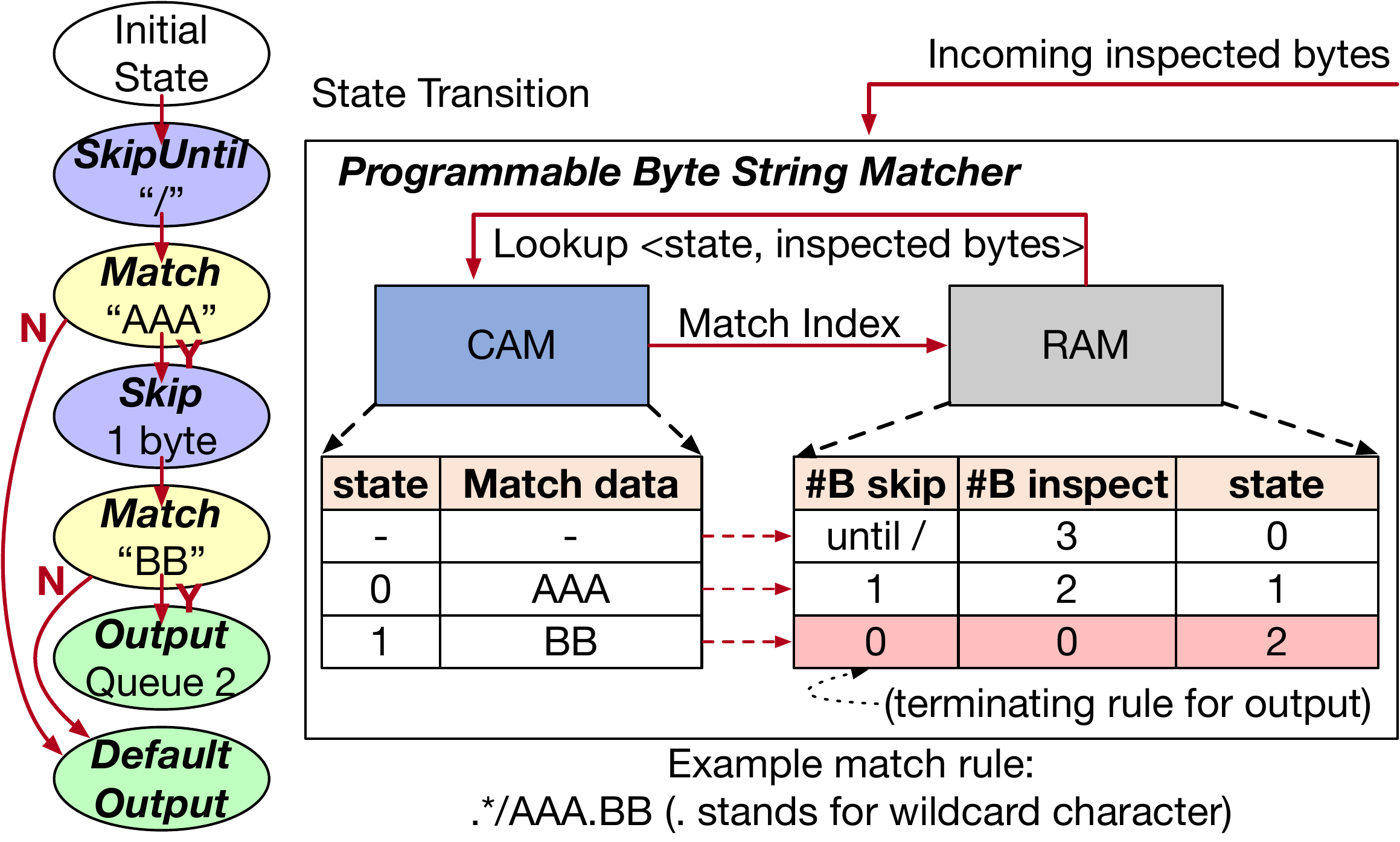}
    \caption{Matcher implements the skip-and-match. Skip and Match are encoded in {\color{black}RAM} and {\color{black}CAM} entries respectively. {\color{black}RAM entries} with skipping and matching 0 bytes denote exit states: it shows the output queue index in the last column.
    \label{fig:rsd}}
\end{figure}

\textit{Matcher.} We take inspiration from implementing hardware packet parsers~\cite{rmt_parser_design} to implement \textit{skip-and-match} in Matcher by mapping \textit{skip}s and \textit{match}es to Random Access Memory (RAM) and Content Addressable Memory (CAM) respectively.
Specifically, as in Figure~\ref{fig:rsd}, a dispatch rule consisting of skip-and-matches can be translated into a state machine, where states alternate between skipping bytes and matching strings until a final output is found. We map states that skip bytes into entries in RAM, and states that match string into entries in a CAM so that the string matching can be implemented as a content-addressable operation. Rule matching then requires going back and forth between these two units, and Matcher outputs an RX queue identifier when a match/mismatch has been found.

Currently, the number of entries for configuring dispatch rules is set to 512 for both RAM and CAM. 
The RAM entry is designed to be 32-bit width: an 8-bit field index to be matched against, two 8-bit numbers of how many bytes to skip and match for one skip-and-match, and an 8-bit state number to represent the translated state machine of corresponding dispatch rule in the Matcher. 
In our implementation, we use the number of skipped bytes set to 0xff to indicate the \textit{SkipUntil} primitive, which means to skip to the character `/'. Otherwise, it is configured as the \textit{Skip} primitive. We ensure the number of skipped bytes in each skip is smaller than 64, as BytePipe can support reading 64B at most simultaneously.

Each CAM entry is 96-bit wide: a 24-bit number to be matched against an 8-bit application ID, 8-bit message type and 8-bit field index, an 8-bit numbered state and an 8-byte (\ie 64-bit) byte string to be matched against.
The terminating rule is encoded into RAM entry of skip-and-match with 0 bytes to skip and match.

\Para{Cache of dispatch result.} We use a small RAM as a stash to cache the dispatching result for each message. Specifically, each stash entry is 41-bit wide: 8-bit cached dispatching result, 32-bit corresponding message ID, and 1-bit indicator of validity. The depth of the stash is set to 128.

\Para{\sysname hardware states.} As described in \S\ref{ssec:qn_hw_arch}, instead of buffering packets for L7 message reassembly on NIC, \sysname only tracks neccesary states for in-flight messages. In our current prototype, the number of tracked message entries is set to 512. The states of each entry consist of a 16-byte descriptor for DMAing packets from NIC to host, a 24-bit timer, and an 8-bit tracker of received individual packets within the message. The timeout for invalidating existing message entries is fixed to 1\si{\second}.

\subsection{\sysname Software Prototype}
\label{ssec:impl_qn_sw}

\begin{figure}[!t]
\centering
\begin{lstlisting}[style=CStyle]
struct qnp_pkt_hdr {
  u8  app_id; // application ID
  u8  msg_type; // indicator of message type
  u32 msg_id; // message ID
  u32 msg_acked_id; // acked message ID
  u8  msg_len_in_pkts; // # of pkt in msg
  u8  pkt_seq_num_in_msg; // pkt's seqnum in msg
  u8  pkt_flag; // indicator of DATA or ACK
  // for optimization
  u8  seg_cnt; // #segments that count for RSD
  u8  paddings[PAD_SIZE]; // for alignment
};
\end{lstlisting}
\caption{QNP packet header definition.\label{fig:qnp_hdr}}
\end{figure}


\begin{figure}[!t]
    \centering
    \begin{subfigure}[b]{\linewidth}
        \centering
        \includegraphics[width=\linewidth]{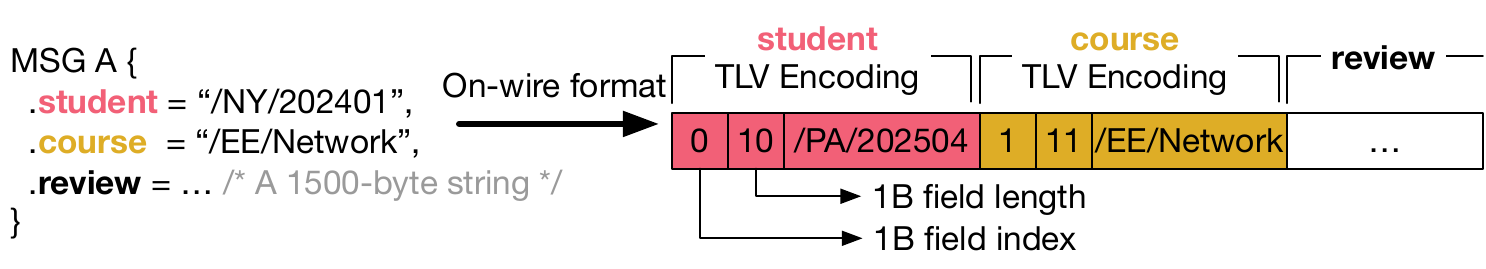}
        \caption{TLV format.\label{fig:qn_onwire_format}}
    \end{subfigure}
    \hfill
    \begin{subfigure}[b]{\linewidth}
        \centering
        \includegraphics[width=\linewidth]{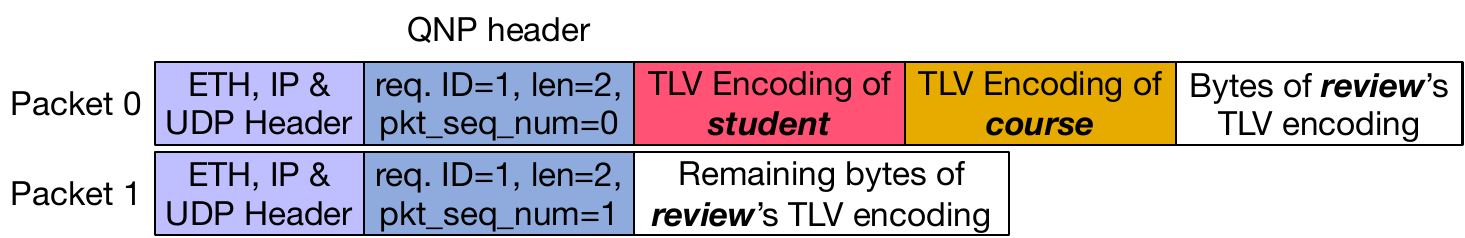}
        \caption{Packet layout.\label{fig:qn_pkt_layout}}
    \end{subfigure}
    \caption{Example of \sysname on-wire format for a message struct with two fields, where fields `student' and `course' are used for dispatch. \label{fig:qn_pkt_format}}
\end{figure}

\Para{\sysname IDL compiler.} \sysname IDL currently offers {\it string} as the only type for the message struct fields---which is enough to represent variable-length byte string---and it assumes no nested message struct definition. We based our \sysname IDL compiler on a protobuf~\cite{protobuf} frontend parser. The \sysname IDL compiler is written in \idlLOC lines of Golang. It takes the message format as input and emits the corresponding data structure and (de)serialization functions in C++.

\Para{\sysname protocol.} For simplicity, instead of directly modifying the complex protocols (\eg TCP~\cite{tcp_protocol}, QUIC~\cite{quic_protocol}), we propose a \sysname protocol (QNP) for \sysname library to easily layout the message packets as desired.
QNP is designed to provide minimum support of reliable delivery by ACKing every message. Primarily, as shown in Figure~\ref{fig:qnp_hdr}, the QNP header consists of three main fields--- a unique message ID, an in-message packet sequence number, and a message length in the number of packets---which allows both sender and receiver to keep track of individual packets of each message.

We implement QNP on top of UDP packets. To simplify hardware implementation, we make sure the packet header (\ie Ethernet, IP, UDP, and QNP) is 512-bit aligned, which is the data width in \sysname hardware. Thus we set {\it PAD\_SIZE} to 8 in Figure~\ref{fig:qnp_hdr}.
QNP's control loop works at the granularity of message, meaning that once an application message is received, \sysname library will piggyback the ACK to the response. Compared with ACKing every individual packet at NIC, it avoids packet amplification that may lead to excessive traffic load on NIC's TX path.

\Para{\sysname on-wire format.} \sysname uses a TLV format---which can effectively express variable-length strings (\eg protobuf~\cite{protobuf}, HTTP/2~\cite{h2_rfc})---to represent on-wire message struct fields in the QNP payload. Take Figure~\ref{fig:qn_onwire_format} as an example, a student field ``/PA/202504'' is encoded as 12 bytes: 1-byte field index, 1-byte field length and 8-byte original string. 
Recall that we also encode a 1-byte field index in the RSD's Matcher configuration, the TLV's field index and length are used by \sysname RSD to ensure dispatch rules are correctly enforced on the target struct fields by matching the field index and check the already parsed field byte length.

\Para{\sysname library and driver.} 
\sysname library implements \sysname APIs (\ie \textit{send\_msg()} and \textit{recv\_msg()}) with QNP. 
Specifically, \textit{send\_msg()} API serializes the message into on-wire bytes using the serialization function generated by \sysname IDL, segments the message into 1500-byte packets, and attaches them with a unique message ID per sender and a corresponding packet sequence number within the message (Figure~\ref{fig:qn_pkt_layout}). Also, fields marked with \textit{dispatch\_info} are guaranteed to be laid out only in the first packets as described in \S\ref{ssec:design_qn_sw}.
These packets of each message are then pushed to the QNP layer, which is responsible for reliable delivery. Once a message is ACKed, the corresponding memories used by its packets are reclaimed at the sender's QNP layer.  
Currently, \sysname assumes each message has at most 4 packets. \textit{recv\_msg()} API grabs the received messages from the QNP layer.

We use DPDK 21.11~\cite{dpdk} to manage packet memory and implement a userspace poll-mode driver to provide kernel-bypassing access to \sysname NIC. In \sysname, the driver is used by the QNP layer to receive the complete messages from the NIC. As discussed in \S\ref{ssec:qn_hw_arch}, \sysname NIC notifies the host of the complete message directly, which is unlike an L7-agnostic NIC that delivers packets to the host. Therefore, for comparison, we also implement a software-based message management in the QNP layer for the baselines using the same userspace driver.

The \sysname library, QNP layer, and driver are written in C++ (\driverLOC LOC).

\subsection{Optimizations of \sysname}
\label{ssec:qn_tput_opt}

\Para{Parallel RSDs.} Since Matcher needs to instruct BytePipe based on the configured dispatch rules, it requires more cycles to dispatch a packet as the complexity (\ie number of skip-and-match) increases. This will bottleneck the NIC processing. 
To alleviate the potential performance bottleneck due to RSD's serial processing, we apply parallel RSDs to process the incoming packets. Messages are sharded across parallel RSDs by their message IDs. A larger number of parallel RSDs improves performance but results in larger hardware resource consumption. Currently, the number of parallel RSDs is set to 4 in \sysname for our FPGA prototype meeting timing constraints at 250 \si{MHz}.

\Para{Explicit indicator of dispatch info length.} Additionally, when Matcher emits the dispatch result, it still needs to wait for BytePipe to finish flushing unused bytes of the same packet, to avoid polluting the processing of the next packet. However, flushing needs cycles. If it takes too long, the temporary packet buffer will be filled up, thus causing a bottleneck in the NIC pipeline.
To this end, we introduce a field \textit{seg\_cnt} in QNP header (Figure~\ref{fig:qnp_hdr}) to indicate how many AXI-S segments, where the segment size is determined by the AXI-S data width (64B in our implementation), are used by Matcher when serializing dispatch\_info fields (\S\ref{ssec:impl_qn_sw}). 
Instead of inserting all the packet AXI-S segments, \textit{seg\_cnt} controls the number of AXI-S segments inserted to RSD's transient packet buffer, thus reducing the cycles for flushing unused bytes.
\section{Evaluation}
\label{sec:evaluation}

\begin{figure}
    \centering
    \includegraphics[width=0.8\linewidth]{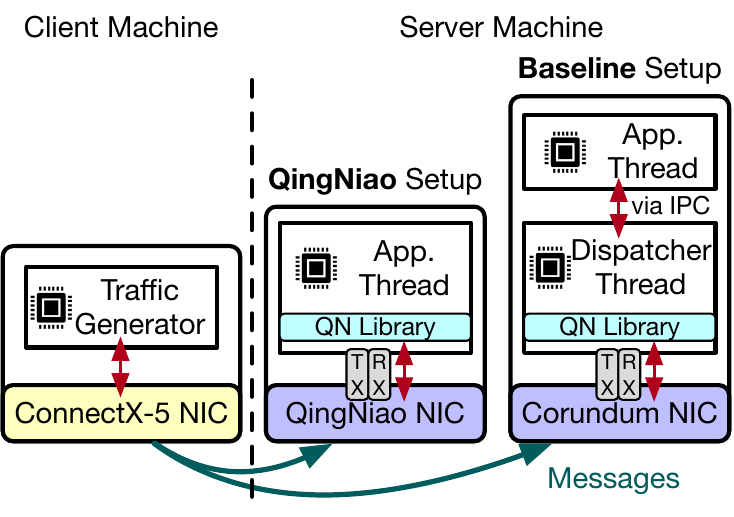}
    \caption{Experimental setup.\label{fig:eval_exp_setup}}
\end{figure}

In this section, we evaluate \sysname using integration with RocksDB~\cite{rocksdb} and conducting microbenchmarks.

\Para{Experimental setup.} As shown in Figure~\ref{fig:eval_exp_setup}, we evaluate \sysname on a server machine with two Intel(R) Xeon(R) Gold 6132 14-core CPUs @ 2.6\si{GHz}, which runs Ubuntu LTS 20.04 with Linux kernel version 5.15.0. Additionally, this server is equipped with an AMD Xilinx Alveo U250 FPGA board~\cite{au250} programmed as a NIC, which is plugged into a PCIe Gen3 x16 slot and is loaded with \sysname hardware prototype and Corundum of the same base commit~\cite{corundum_base_commit} for evaluating baselines (\S\ref{ssec:eval_rocksdb}). 
Our client machine is equipped with an Intel(R) Core(TM) i7-9700 8-core CPU @ 3.0\si{GHz} and a Nvidia Mellanox CX-5 100GbE NIC~\cite{mlnx_cx5}.
We disable hyper-threading on both machines and run processes within the same NUMA domain of the NIC as cross-NUMA communication is expensive. The NIC-to-NIC round-trip latency between two machines is $\sim$800 \si{\ns} in our testbed.

\Para{Baselines.}
For a direct and fair comparison,\footnote{The performance numbers of our baselines are much better than our tested numbers with Envoy (\S\ref{sec:background}).} we build software RSD dispatchers atop Corundum~\cite{corundum} using RSS and IPC mechanisms to allow messages to be exchanged between the software dispatcher and application threads.\footnote{The software dispatchers implement a straw-man match against the configured rules in a for loop in each skip-and-match.} The IPC mechanisms between the dispatcher and application threads are (1) DPDK RTE Ring~\cite{dpdk}; (2) a state-of-the-art (\ie SPRIGHT~\cite{spright}) approach that leverages eBPF; (3) UNIX Domain Socket (UDS). 

As in Figure~\ref{fig:eval_exp_setup}, we use a shared-nothing architecture: when one thread (\eg application, software dispatcher)---that interacts with the NIC---is launched, one free TX/RX queue pair is exclusively initialized and allocated to it. 
Throughout the evaluation, each thread is pinned on a unique core. All the applications and the software dispatchers are implemented in \appLOC LOC of C++.

\Para{Methodology.}  
In \S\ref{ssec:eval_rocksdb}, we integrate \sysname with RocksDB~\cite{rocksdb}, and compare it with the software dispatcher with state-of-the-art IPC mechanisms as baselinese. We demonstrate that \sysname can (1) provide programmability to support dispatching rules; (2) benefit applications by improving throughput while decreasing latency.
In \S\ref{ssec:eval_microbench}, we conduct microbenchmarks to show insights: (1) \sysname achieves a similar performance compared to L7-agnostic hardware dispatchers (\ie RingLeader~\cite{ringleader} and RSS~\cite{rss}); (2) \sysname outperforms the corresponding software implementations of L7 dispatch. We also show \sysname's hardware resource consumption.

\Para{Configurations of dispatch rules.} As in \S\ref{ssec:impl_qn_hw}, the total number of dispatch rules is constrained by the number of entries of RAM \& CAM in RSD, which is 512 in our prototype. To show \sysname's support of programmability, we explore 3 dimensions of \sysname's skip-and-match abstraction: (1) the number of skip-and-matches; (2) the number of dispatch rules; (3) the number of matched bytes per skip-and-match,\footnote{Results are presented in Appendix~\ref{sec:appendix_exp_on_bytes}.} where larger numbers mean that rules are more complex. Note that in skip-and-match (\S\ref{sssec:qn_rsd}), Skip primitive is designed so that a set of dispatch rules are matched against while SkipUntil is used to skip to the specified character. To show how \sysname performs under more complex rules, we fix the number of SkipUntil to 1 and vary the number of Skips in skip-and-match. When testing, we vary one dimension and keep the other two fixed.\footnote{The numbers of skip-and-matches (\ie Skips), dispatch rules, and matched bytes per skip-and-match are fixed to 4, 32, and 5 respectively, when they are not the varying dimension.} For each skip-and-match, we randomly generate a set of rules (\eg 32) to be matched against. 
Throughout the evaluation, we guarantee the configurations do not deplete the 512 entries of both RAM and CAM in RSD.
When dispatching, we ensure that every generated message goes through all the configured number of skip-and-match, and hits exactly one dispatch rule.

To generate workloads, on the client machine, since we have to ensure messages are reliably delivered to test the maximal throughput, we developed a closed-loop traffic generator that generates messages based on DPDK v21.04~\cite{dpdk}. It uses 4 cores to send and receive messages. When testing with RSS, we ensure incoming traffic is evenly distributed to each core. Also, we guarantee the traffic generator is not the bottleneck in any conducted experiments.

\subsection{Integration with RocksDB}
\label{ssec:eval_rocksdb}

\begin{figure*}[!t]
    \begin{minipage}[t]{0.48\textwidth}
        \begin{subfigure}[b]{0.48\linewidth}
            \includegraphics[width=\linewidth]{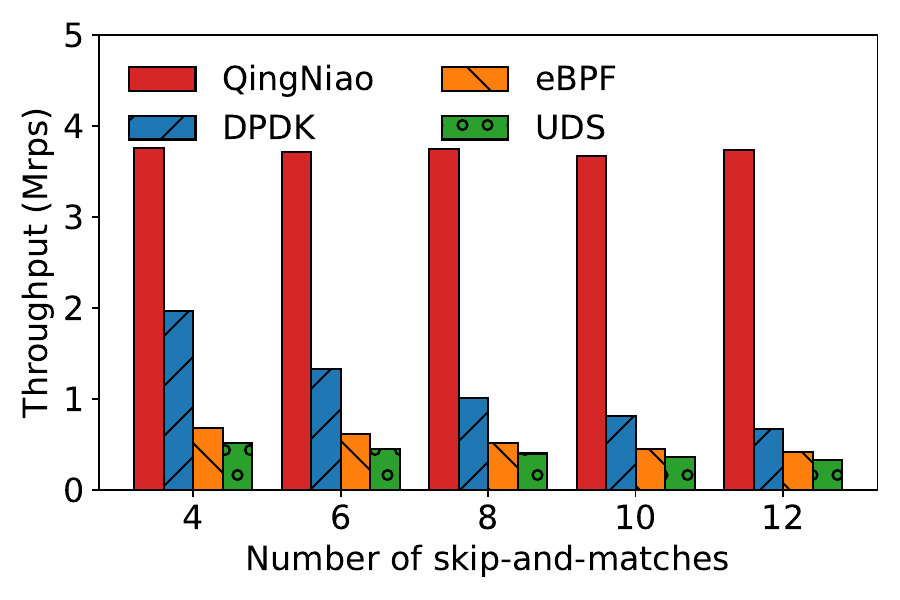}
            \caption{Throughput comparison.\label{expfig:qn_rocksdb_skip_tput}}
        \end{subfigure}
        \hfill
        \begin{subfigure}[b]{0.48\linewidth}
            \includegraphics[width=\linewidth]{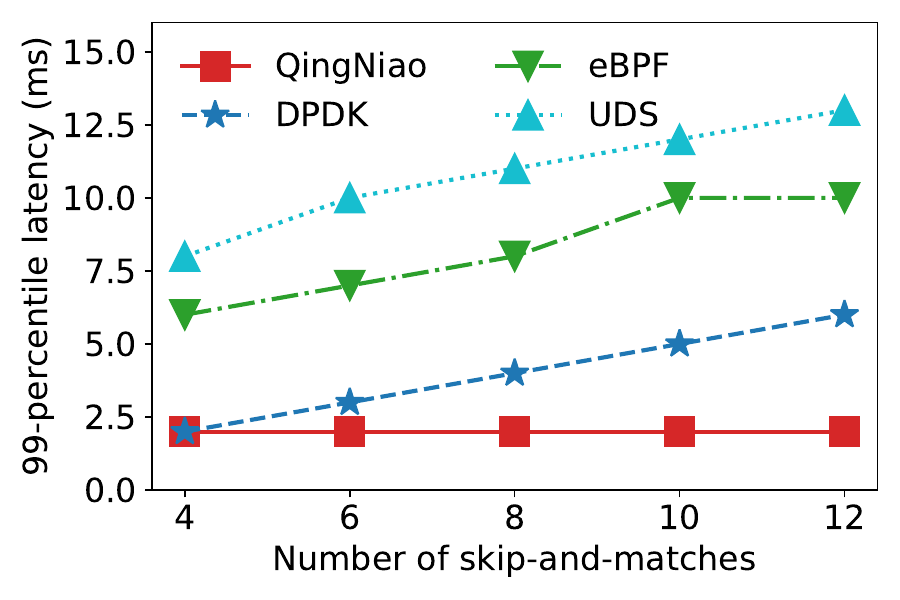}
            \caption{Latency comparison.\label{expfig:qn_rocksdb_skip_latency}}
        \end{subfigure}
        \caption{On throughput, \sysname outperforms DPDK dispatcher by 3.69$\times$ averagely with varying \# of skip-and-matches. \label{expfig:rocksdb_skip}}
    \end{minipage}
    \hfill
    \begin{minipage}[t]{0.48\textwidth}
        \begin{subfigure}[b]{0.48\linewidth}
            \includegraphics[width=\linewidth]{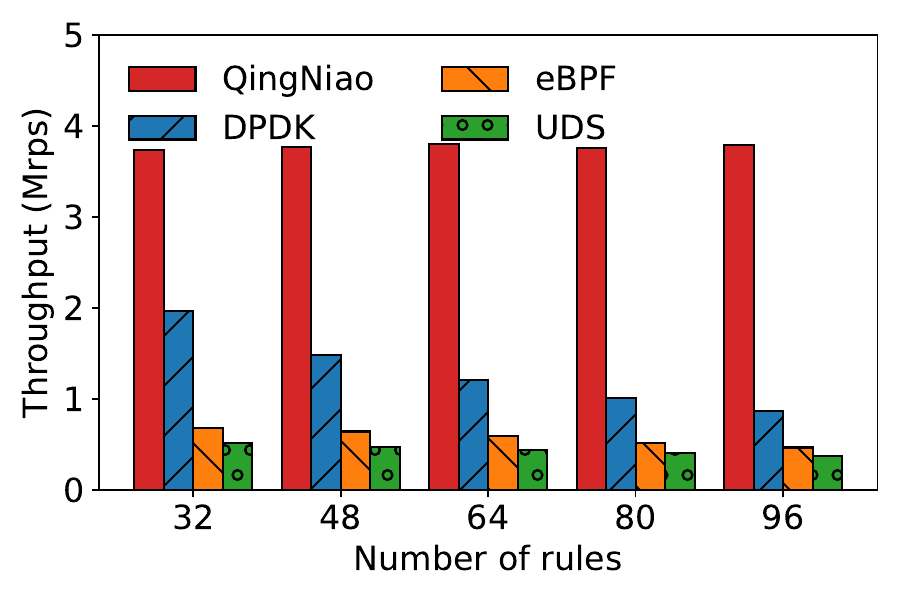}
            \caption{Throughput comparison.\label{expfig:qn_rocksdb_rule_tput}}
        \end{subfigure}
        \hfill
        \begin{subfigure}[b]{0.48\linewidth}
            \includegraphics[width=\linewidth]{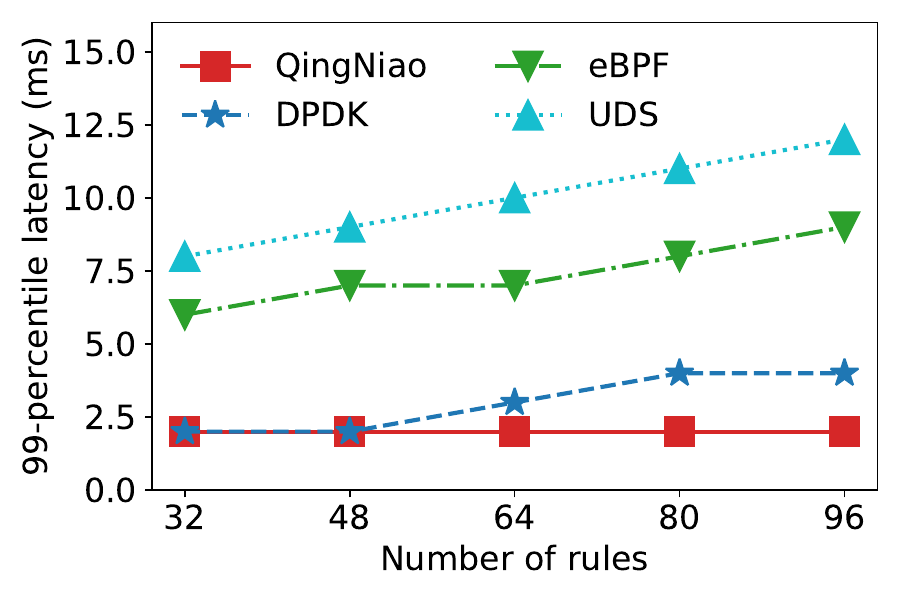}
            \caption{Latency comparison.\label{expfig:qn_rocksdb_rule_latency}}
        \end{subfigure}
        \caption{On throughput, \sysname outperforms DPDK dispatcher by 3.12$\times$ averagely with varying \# of rules dimension. \label{expfig:rocksdb_rule}}
    \end{minipage}
\end{figure*}

\Para{Workloads.} We choose RocksDB~\cite{rocksdb} v7.9.2, a wide-deployed in-memory key-value store, as our application service to serve GET messages. We configure RocksDB to be backed by a 4G {\it tmpfs} folder. It pre-loads 3.14 million 62-byte keys with 64-byte values. Keys are evenly partitioned into 48 ranges, which are then equally assigned to each application thread. From the traffic generator, within each partition, message keys follow a Zipfian distribution with a Zipfian parameter equal to 0.9. The dispatch rules evenly map the keys of messages to the configured number of application threads.
We test the end-to-end throughput and latency, where numbers are reported from the traffic generator and averaged for 5 runs.

\Para{Setup.} 
As our CPU only has 14 cores on a single NUMA node, for the baselines, we fix the number of dispatchers to 4, and the number of application threads to 8. This experimental setting gives the best performance for all the baselines in our test. For comparison, \sysname uses the same amount of CPU cores, which is 12.

Figures~\ref{expfig:rocksdb_skip} and~\ref{expfig:rocksdb_rule} demonstrate that \sysname is able to offload varied dispatch rules efficiently. 
In terms of achieved throughput, as we can observe from Figure~\ref{expfig:qn_rocksdb_skip_tput} and Figure~\ref{expfig:qn_rocksdb_rule_tput}, \sysname outperforms DPDK, eBPF and UDS by 3.69$\times$, 7.15$\times$, 9.22$\times$ and 3.12$\times$, 6.6$\times$, 8.64$\times$, respectively. The more complex dispatch rules are, the more performance gain is achieved by \sysname. This is because CAM's inherent feature allows it to match against multiple (\eg tens in our case) entries in a single cycle---unlike software dispatch. 

In terms of end-to-end latency, as shown in Figure~\ref{expfig:qn_rocksdb_skip_latency} and~\ref{expfig:qn_rocksdb_rule_latency}, \sysname reduces the 99-percentile latency by 42\%, 74.62\%, 80.95\% and 26.67\%, 72.46\%, 79.59\%, respectively. When testing with simpler dispatch rules, software dispatch may be quick as CPU frequency is 12 times the FPGA frequency, which also explains why DPDK can achieve similar latency as \sysname in such cases.

\Para{Takeaway \#1.} \sysname provides programmability to support expressive dispatch rules efficiently.
By offloading L7 dispatch and integrating with \sysname, RocksDB application outperforms its configuration of the software dispatcher with state-of-the-art IPC mechanism (\ie eBPF), by achieving $\sim$7$\times$ throughput and reducing the 99-percentile latency by $\sim$74\%, as \sysname takes advantage of parallel matching in hardware and avoids the extra hop of IPC.

\subsection{Microbenchmarks}
\label{ssec:eval_microbench}

Throughout the microbenchmarks below, unless stated, the traffic generator generates 128-byte messages, including all the packet headers (\S\ref{ssec:impl_qn_hw}), and waits for a 128-byte response from the server application thread. We term this application as PingPong, where each PingPong application has its unique message format and dispatch rules. 

In the following, we have two scenarios: (1) \textit{single} PingPong, where it compares \sysname with L7-agnostic hardware dispatchers and \sysname's software implementation; (2) \textit{multiple} PingPong, where it demonstrates \sysname's support of multiple concurrent applications.
Numbers are reported on the traffic generator and averaged for 5 runs.

\subsubsection{Single PingPong Application}
\label{sssec:mb_single}

This section evaluates \sysname using a \textit{single} PingPong.

\begin{table}[!t]
\begin{center}
    \scriptsize
    \begin{tabular}{L{0.5in}C{0.08in}C{0.08in}C{0.08in}C{0.08in}C{0.08in}C{0.08in}C{0.08in}C{0.15in}C{0.4in}}
        \toprule
        \multicolumn{1}{c}{\textbf{System}} & \multicolumn{7}{c}{\textbf{\sysname}} & \multicolumn{1}{c}{\textbf{RSS}} & \multicolumn{1}{c}{\textbf{RingLeader}} \\
        \midrule
        \textbf{\# of S\&M} & 1 & 2 & 4 & 8 & 16 & 32 & 48 & - & - \\
        \midrule
        \textbf{\# of Cycles} & 15 & 21 & 33 & 57 & 105 & 201 & 297 & 3 & 25 \\
        \midrule
        \textbf{Latency (\si{\nano\second})} & 448 & 469 & 517 & 613 & 805 & 1189 & 1573 & - & - \\
        \bottomrule
    \end{tabular}
\caption{\sysname's latency reported in simulation and measured on testbed. \sysname adds 6 cycles per skip-and-match (S\&M).\label{tab:qn_latency}}
\end{center}
\end{table}

\Para{Comparison to state-of-the-art L3/4 hardware dispatchers.} In terms of the {\it latency} of dispatching, L3/4 hardware dispatchers have a fixed number of cycles, since they are processing fixed-length fields (\eg IP addresses) at known offsets. In comparison, \sysname's processing latency is dependent on the number of skip-and-matches. To show this, we report the exact number of cycles in the simulation, which is cycle-accurate. Additionally, we timestamp the received packet at both the Ethernet MAC+PHY and the RSD modules (Figure~\ref{fig:qn_nic_arch}) to measure the ingress latency---including all ingress processing (\eg packet filter, etc.)---on the testbed. RSS used in Corundum~\cite{corundum} computes a hash value using Toeplitz hash algorithm~\cite{toeplitz_hash} based on packet header's 5-tuple fields (\ie IP protocol, source/destination IP addresses, and source/destination ports). We send 4\si{K} messages, with a gap of 50\si{\milli\second} between messages to ensure there is no message queuing. Measured numbers are reported on average.

Compared with L3/4 dispatcher RSS and RingLeader~\cite{ringleader}\footnote{Test of RingLeader is based on its repository~\cite{ringleader_repo} with default settings and VFIO~\cite{vfio} disabled. We modify it to support QNP closed-loop test.} that take 3 and 25 cycles respectively to compute a dispatch decision, \sysname incurs increasing latency as L7 matching becomes more complex as shown in Table~\ref{tab:qn_latency}. It reaches 297 cycles and leads to a total of 1573 \si{\ns} ingress latency for 48 skip-and-matches. The latency can be optimized once prototyped using ASIC,\footnote{With the same parameters in \S\ref{ssec:impl_qn_hw}, we use the Synopsys DC~\cite{synopsys_dc} and FreePDK45nm technology library~\cite{freepdk45} to synthesize \sysname with 4 parallel RSDs. It reports that \sysname can run at 1\si{GHz} with 0.766\si{\milli\meter\squared} chip area.} since our FPGA prototype can only run at 250\si{MHz} with 4-\si{ns} cycle.

\begin{figure*}[!t]
    \begin{minipage}[t]{0.32\linewidth}
        \centering
        \includegraphics[width=\linewidth]{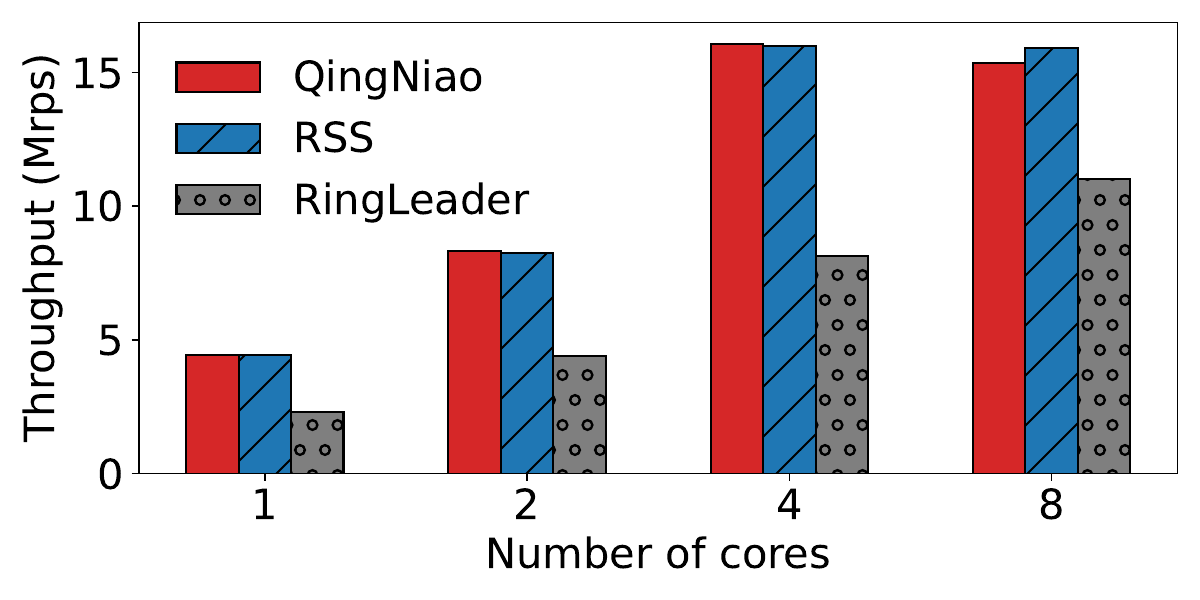}
        \caption{\sysname achieves comparable performance as hardware L3/4 dispatcher.\label{expfig:qn_vs_hw_l34}}
    \end{minipage}
    \hfill
    \begin{minipage}[t]{0.32\linewidth}
        \centering
        \includegraphics[width=\linewidth]{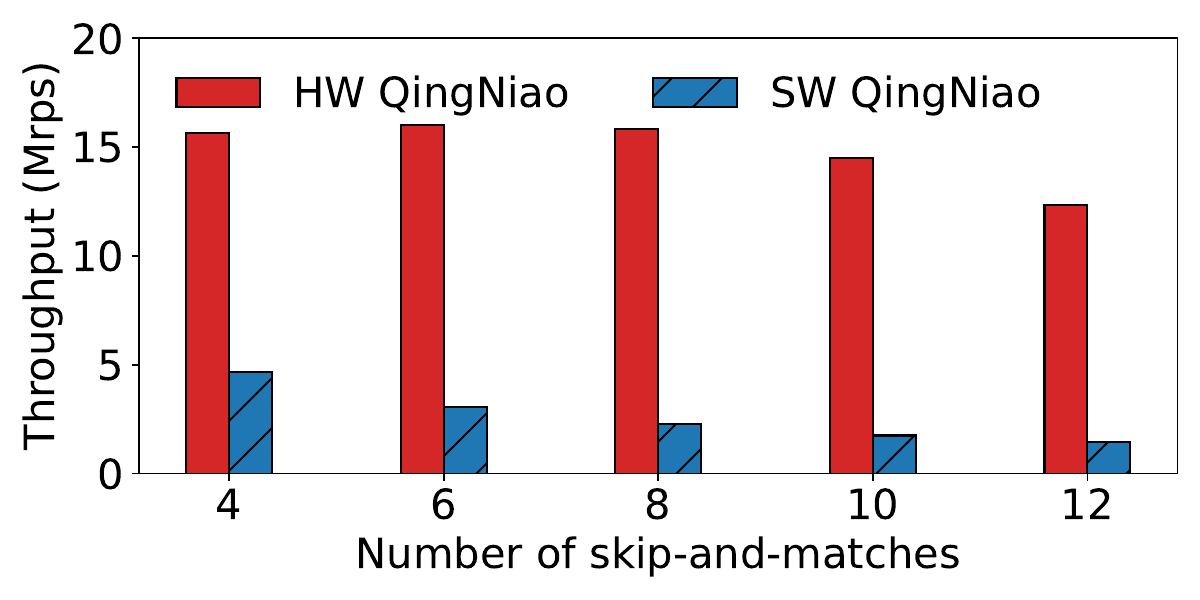}
        \caption{\sysname outperforms software implementation by 6.49$\times$ averagely with varying \# of skip-and-matches.\label{expfig:qn_hw_sw_skips}}
    \end{minipage}
    \hfill
    \begin{minipage}[t]{0.32\linewidth}
        \centering
        \includegraphics[width=\linewidth]{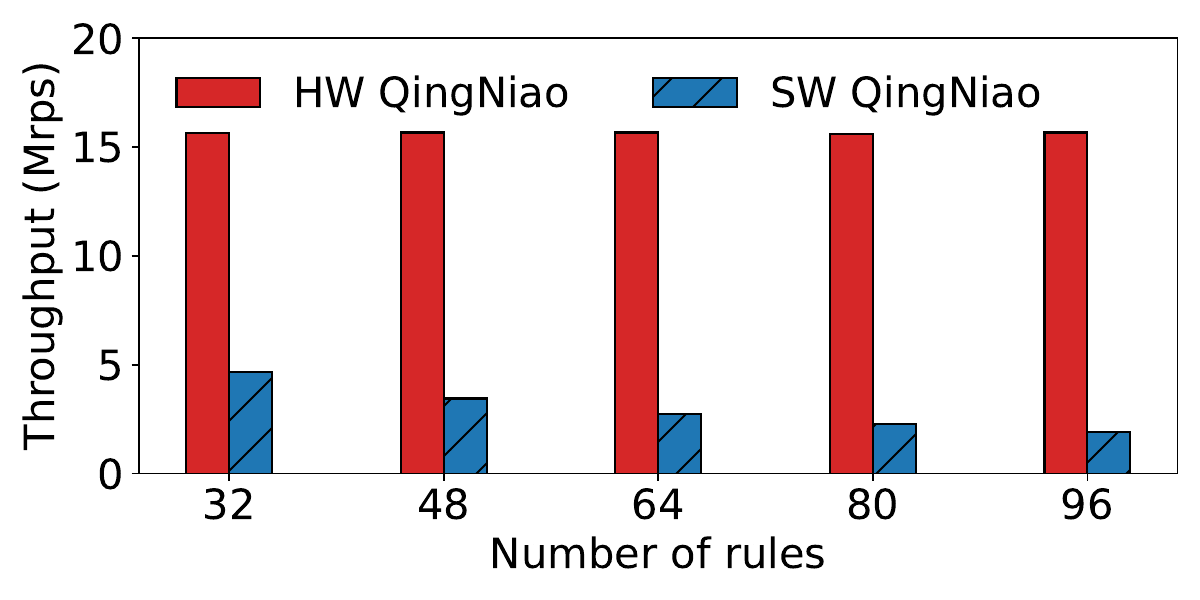}
        \caption{\sysname outperforms software implementation by 5.74$\times$ averagely with varying \# of rules.\label{expfig:qn_hw_sw_rules}}
    \end{minipage}
\end{figure*}

On the aspect of {\it throughput}, we compare \sysname with RSS and RingLeader~\cite{ringleader}. To show its maximum capability and scalability, we vary the number of cores (\ie number of queue pairs, application threads) and fix the number of skip-and-match to 1. 
As observed in Figure~\ref{expfig:qn_vs_hw_l34},\footnote{RingLeader's performance numbers reported on our testbed are lower than the numbers originally reported in~\cite{ringleader}. One potential explanation is that we are using different hardware setups.} despite the incurred additional processing latency as we described, \sysname achieves comparable throughput with RSS and can get 16.06\si{M}rps for 4 cores. However, \sysname's performance drops slightly to 15.34\si{M}rps for 8 cores. This can be the result of cache contention as the number of cores in the same NUMA domain increases. 

\Para{Comparison to \sysname's software implementation.} To demonstrate \sysname's effectiveness of L7 dispatch offload, we additionally build software RSD: when a thread receives a message from Corundum NIC, it simply runs the software RSD algorithm and then echoes the response back without dispatching (\ie no IPC involved). The number of cores is fixed to 8 in this set of experiments.
As we can see from Figure~\ref{expfig:qn_hw_sw_skips} and~\ref{expfig:qn_hw_sw_rules}, \sysname outperforms its software implementation by 6.49$\times$ and 5.74$\times$, respectively. However, \sysname performance also drops as the number of skip-and-match increases. This is because the processing bottleneck shifts to RSD as we describe below.

\begin{figure*}[!t]
    \begin{minipage}[b]{0.32\linewidth}
        \centering
        \includegraphics[width=\linewidth]{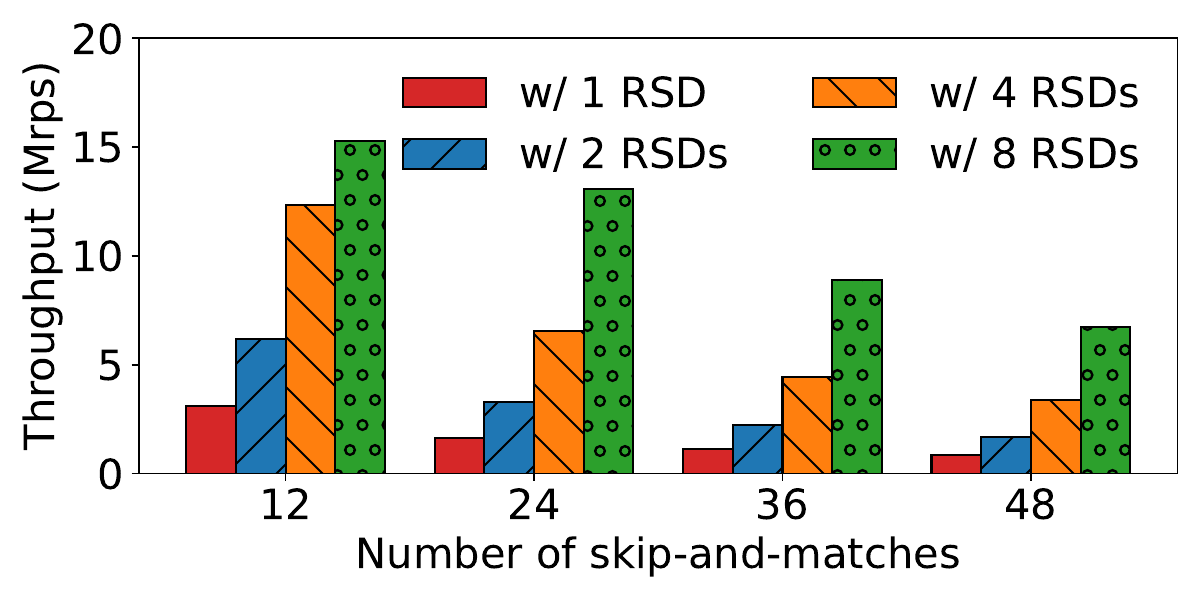}
        \caption{Having parallel RSDs improves throughput.\label{expfig:qn_multi_rsds}}
    \end{minipage}
    \hfill
    \begin{minipage}[b]{0.32\linewidth}
        \centering
        \includegraphics[width=\linewidth]{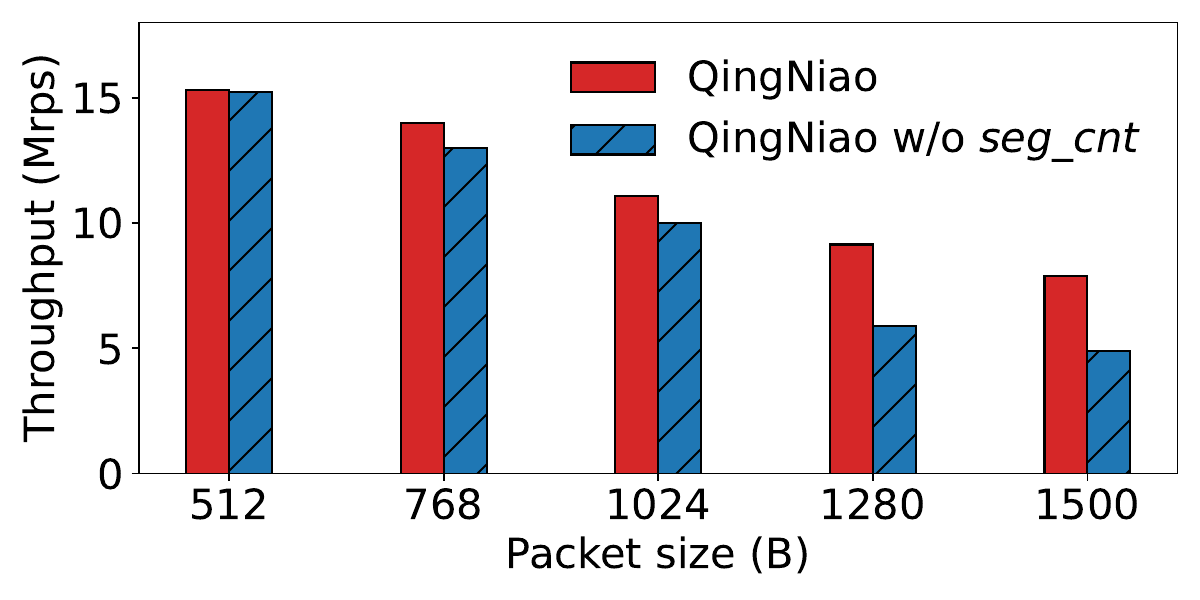}
        \caption{Introducing seg\_cnt improves throughput.\label{expfig:qn_seg_cnt}}
    \end{minipage}
    \hfill
    \begin{minipage}[b]{0.32\linewidth}
        \centering
        \includegraphics[width=\linewidth]{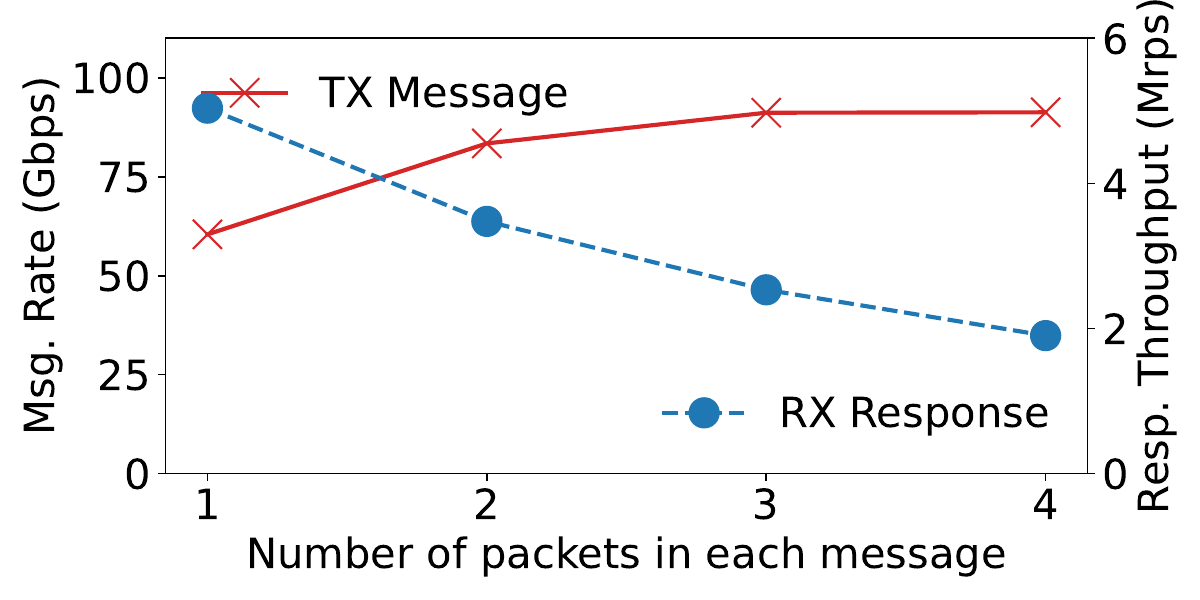}
        \caption{\sysname supports multi-packet messages.\label{expfig:qn_multi_pkt_msgs}}
    \end{minipage}
\end{figure*}

\begin{table}[!t]
\begin{center}
    \scriptsize
    \begin{tabular}{crrr}
        \toprule
        \textbf{\# of parallel RSDs} & \multicolumn{1}{c}{\textbf{LUTs as logic}} & \multicolumn{1}{c}{\textbf{LUTs as memory}} & \multicolumn{1}{c}{\textbf{BRAM}} \\
        \midrule 
        1           & 20033 (1.16\%)     & 2644 (0.334\%)    &  3 (0.112\%) \\
        2           & 40098 (2.32\%)     & 5288 (0.668\%)    &  6 (0.223\%) \\
        4           & 81589 (4.72\%)     & 10576 (1.337\%)   &  12 (0.446\%) \\
        8           & 163959 (9.45\%)    & 21152 (2.674\%)   &  24 (0.893\%) \\
        \bottomrule
    \end{tabular}
\caption{FPGA resources usage for module of parallel RSDs.}
\label{tab:rsd_resource_usage}
\end{center}
\end{table}

\Para{Effectiveness of \sysname introduced optimizations (\S\ref{ssec:impl_qn_hw})} As discussed, when the number of skip-and-match increases, \sysname's RSD needs more cycles to process one single packet, which stalls the entire NIC pipeline. Having multiple parallel RSDs alleviates this problem. As shown in Figure~\ref{expfig:qn_multi_rsds}, the performance of \sysname with 8 parallel RSDs doubles as compared to \sysname configured with 4 parallel RSDs when the bottleneck shifts from DMA to RSD as the number of skip-and-matches increases from 12 to 24. However, there is a tradeoff between performance and resource consumption. As shown in Table~\ref{tab:rsd_resource_usage}, the FPGA memory usage doubles as we double the number of RSDs.

Besides, introducing {\it seg\_cnt} field in the QNP header also helps improve \sysname performance as shown in Figure~\ref{expfig:qn_seg_cnt}. The intuition behind this is that introducing {\it seg\_cnt} limits the number of bytes buffered in BytePipe, thus RSD can emit dispatch decisions more quickly, which allows RSD to sustain higher rates.

\Para{\sysname supports multi-packet message.} As in Figure~\ref{expfig:qn_multi_pkt_msgs}, with 1 core, \sysname achieves 91.28 Gbps for 4-packet messages where each packet chunk size is 1500 byte, which reaches the limit of our current implementation of the QNP stack. Accessing cached dispatch results only needs 2 cycles.

\Para{Takeaway \#2.} \sysname achieves similar performance compared to L3/4 hardware dispatchers (\ie RSS and RingLeader~\cite{ringleader}) and outperforms its corresponding software implementation as evaluated. However, since \sysname's processing time is dependent on the number of skip-and-matches in each rule, RSD can be a potential bottleneck (\S\ref{ssec:impl_qn_hw}). We show that the proposed optimization techniques (\S\ref{ssec:impl_qn_hw}) in \sysname can alleviate this problem.

\subsubsection{Multiple PingPong Applications}
\label{sssec:mb_multi}

To show \sysname's ability to simultaneously support multiple applications, we use two different messages matching different rules to emulate \textit{multiple} PingPong applications.

\begin{figure}[!t]
    \begin{minipage}[t]{0.48\linewidth}
        \centering
        \includegraphics[width=\linewidth]{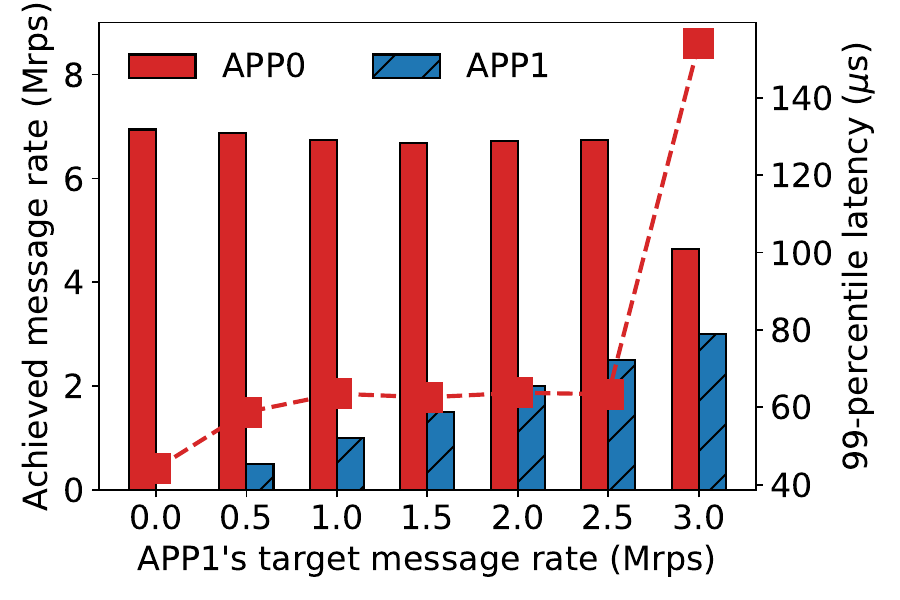}
        \caption{APP0's throughput (red bar) decreases and latency (red curve) increases as APP1's rate increases.\label{expfig:qn_multi_app_tput}}
    \end{minipage}
    \hfill
    \begin{minipage}[t]{0.48\linewidth}
        \centering
        \includegraphics[width=\linewidth]{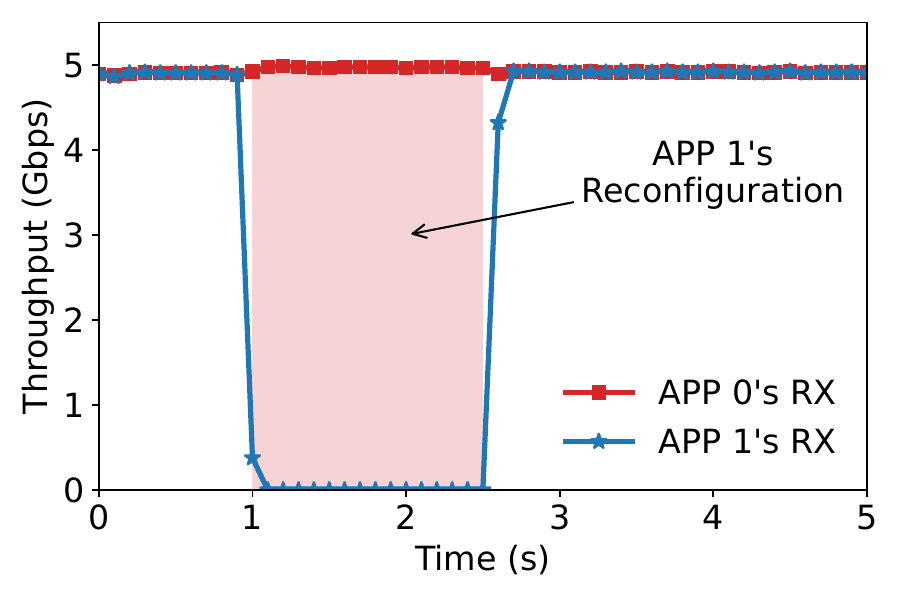}
        \caption{APP0's rate is not impacted when APP1 is reconfiguring in \sysname at runtime.\label{expfig:disruption_free_reconf}}
    \end{minipage}
\end{figure}

\Para{Multiple applications with varied message formats.} As we have shown in \S\ref{sssec:mb_single}, \sysname processing latency increases as the number of skip-and-matches increases. To show the impact of collocating messages with different processing latencies, we configure two different dispatch rules: we set the number of skip-and-matches to 2 (APP0) and 48 (APP1).

Running them separately gives us a throughput of 15.7\si{M}rps and 3.37\si{M}rps, which are bottlenecked at DMA and RSD respectively. To leave enough DMA processing capability, we fix APP0's target rate to 7\si{M}rps and gradually increase APP1's target rate. As in Figure~\ref{expfig:qn_multi_app_tput}, when increasing APP1's rate from 0 to 2.5\si{M}rps, APP0's achieved throughput decreases slightly and latency increases. This is because \sysname shards incoming messages into different RSDs by hashing message IDs, which delays APP0's processing. As APP1's rate reaches 3\si{M}rps from 2.5\si{M}rps, APP0's latency increases by 1.43$\times$, since messages start to queue up at RSD processing. This is the result of \sysname's design choice of being work-conserving rather than providing strong isolation between different messages.

\Para{Support of disruption-free reconfiguration.} We configure two applications to send messages at 5\si{Gbps} constantly. Individual response rates are measured at traffic generator with 0.1 \si{s} granularity. As Figure~\ref{expfig:disruption_free_reconf} shows, APP 1's rate drops to 0 when it starts to reconfigure at 1\si{s}, while APP 0's rate is still maintained at 5 \si{Gbps}. After this 1.5-\si{s} reconfiguration, APP 1's rate recovers to 5 \si{Gbps}, which demonstrates \sysname's support of disruption-free reconfiguration at runtime.

\Para{Takeaway \#3.} As shown, \sysname supports disruption-free reconfiguration but does not provide isolation mechanisms among traffic. Simultaneously serving messages of different formats in \sysname may impact each other as described. We discuss how to provide isolation guarantees in \S\ref{sec:discussion}.

\subsubsection{Resource Usage of \sysname}

\begin{table}[!t]
\begin{center}
    \scriptsize
    \begin{tabular}{lrrr}
        \toprule
        \multicolumn{1}{c}{\textbf{HW Resources}} & \multicolumn{1}{c}{\textbf{\sysname}} & \multicolumn{1}{c}{\textbf{Corundum}} & \multicolumn{1}{c}{\textbf{Additional Usage}} \\
        \midrule 
        LUT             & 222622 (12.88\%)       & 53723 (3.11\%)  & 314.4\% \\
        LUTRAM          & 65038 (8.22\%)         & 10978 (1.39\%)  & 492.4\% \\
        BRAM            & 196.5 (7.31\%)           & 178 (6.62\%)    & 10.4\% \\
        URAM            & 10 (0.78\%)            & 10 (0.78\%)     & 0\% \\
        \bottomrule
    \end{tabular}
\caption{FPGA resource usage of \sysname vs Corundum.}
\label{tab:qn_resource_usage}
\end{center}
\end{table}

As shown in Table~\ref{tab:qn_resource_usage}, \sysname uses more FPGA resources compared to the base Corundum NIC: \sysname uses about 4.92$\times$ more LUTRAM and 10.4\% more BRAM than Corundum. Specifically, this additional resource usage mainly comes from implementing RSD, RAM, and CAM entries to store dispatch rules in \sysname.

\section{Discussion}
\label{sec:discussion}

\Para{\sysname's limitations.} As a research prototype, \sysname has two design limitations:
First, as shown, a message's processing latency is dependent on the complexity of dispatch rules in RSD. We have not yet designed any isolation mechanisms to ensure traffic is isolated when competing for processing bandwidth. Existing works can be applied: (1) on hardware, we can enforce rate limiting by assigning PIFO~\cite{pifo} ranks to applications; (2) on end-host, we can enforce end-to-end isolation policies by employing a throughput probe like Yama~\cite{socc_yama}. 
Second, \sysname hardware requires that only the first packets contain dispatch information, which limits the types of application messages that can be used, requiring that the per-message dispatch information is limited to an MTU.

\Para{How do \sysname's lessons generalize?} 
\sysname changes how application messages are encoded and segmented to offload L7 dispatch to NICs, in particular requiring that the application message's first packet contain all fields that can be used for dispatch. Encoding and segmenting messages in this way means that we do not need to implement reassembly logic in the NIC, nor buffer packets in the NIC. 

However, one might worry that this encoding makes it hard to port \sysname's design to transport protocols like TCP~\cite{tcp_protocol} and QUIC~\cite{quic_protocol} which are stream-oriented. Using \sysname with TCP or QUIC is feasible, though it might require minor implementation changes on the sender side. Note, no protocol changes are required.

Specifically, the sender's protocol implementation would need to change in two ways (a) it must allow the application to packet align each application, \ie ensure that each packet contains data from a single message; and (b) allow the application to determine how much (application) data is sent in each packet. Once this is done, implementing \sysname on top of this modified implementation merely requires changing the application message format to ensure that a message ID appears in each packet, and that the dispatch information appears in the first packet. 

Similarly, \sysname can also be easily applied to message-oriented protocols such as Homa~\cite{homa} and MTP~\cite{mtp}, where again only payloads need to change to accommodate our encoding.

\Para{End-to-end Encryption and \sysname?} 
End-to-end encryption (as provided by TLS~\cite{tls}) can be a challenge for \sysname, since it needs to operate on application data. 
We can adopt approaches like Google PSP~\cite{google_psp} if end-to-end encryption is desired. This approach would require administrators to use an out-of-band mechanism to configure (and update) cipher keys, but allow the NIC to decrypt packets before they are processed by \sysname. Thus, it is feasible to deploy \sysname-like approaches in environments where end-to-end encryption is necessary.

\section{Related Work}
\label{sec:related_work}

\Para{Message dispatching and scheduling.} Recent works~\cite{caladan, shenango, persephone, ringleader, shinjuku} design software dispatching and scheduling mechanisms to better utilize CPUs. Dispatching is inherently orthogonal to scheduling: dispatching decides where the messages should go while scheduling determines in which order the messages should be served. MICA~\cite{mica} dispatches messages using a client-assisted technique to reduce L7 dispatch overhead. However, it assumes single-packet messages and requires exposing underlying server configurations to the clients, which brings management concerns.
Similar to~\cite{ringleader}, \sysname also offloads dispatch to hardware to save CPUs but \sysname targets L7 dispatch.

\Para{Hardware acceleration for RPCs.} Prior work from Google~\cite{google_profiling} showed that a significant fraction of datacenter CPU cycles are spent processing RPC messages. Consequently, there have been several efforts to offload portions of this processing to specialized hardware, but focus on different problems than us. 

Protoacc~\cite{protoacc}, OptimusPrime~\cite{optimus_prime}, and Cereal~\cite{cereal} developed hardware offloads for serialization and deserialization, which is a common component of these overheads. These works are complementary to ours.

One work that addresses the problem of offloading ``dispatch'' is Cerebros~\cite{cerebros}. However, they address a \emph{different} type of dispatch: \textbf{code dispatch}, where the hardware calls a function to handle a received RPC message. By contrast, we address the \textbf{process dispatch} problem, i.e., we decide what process should receive a message. These problems are useful in different places: code dispatch is required when implementing RPC within a process, while process dispatch is used to shard messages across multiple instances of the same program. These problems also require different solutions: Cerebros (and software code dispatch approaches) use a table that maps RPC type to function, while \name (and process dispatch solutions) match message contents to decide where to forward the message.

\section{Conclusion}
\label{sec:conclusion}
We began this project hoping to reduce the overheads of software L7 dispatch by offloading it to NICs.
Our initial plan was to focus solely on the NIC design, but we soon ran into a roadblock because of current approaches for encoding multi-packet application messages, and limitations of how matching policies are specified and implemented in hardware.
Consequently, we had to rethink our approach: \sysname adopts a holistic design that uses a custom-made application message encoding that allows the hardware to process individual packets (rather than needing access to the whole message), an abstraction that is rich enough to specify policies over variable length fields efficiently implementable in hardware, and a software library that minimizes developer effort required to adopt \sysname. We have prototyped \sysname on an FPGA, demonstrating its viability. Our code is available at \arxivrepo.

\clearpage
\balance
\bibliographystyle{abbrv}
\bibliography{main}

\clearpage
\nobalance
\appendix
\section{Experiment on varying \# of matched bytes}
\label{sec:appendix_exp_on_bytes}

\begin{figure}[!t]
    \centering
    \includegraphics[width=0.8\linewidth]{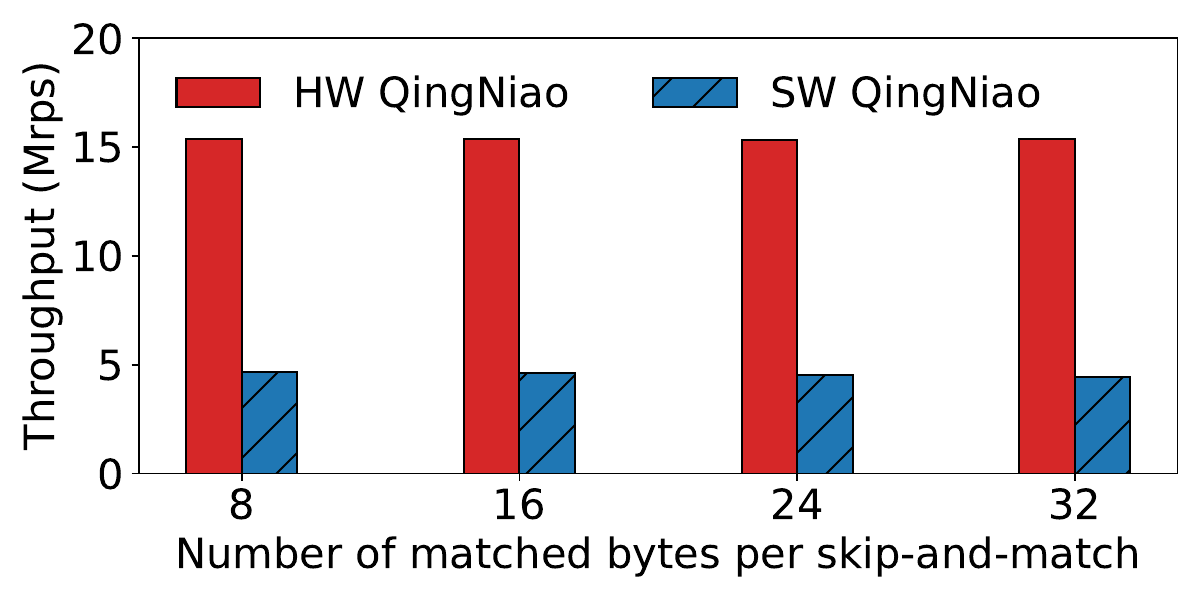}
    \caption{\sysname outperforms its software implementation by 3.36$\times$ averagely with varying \# of matched bytes.\label{expfig:qn_hw_sw_chars}}
\end{figure}

To show how the third dimension in \S\ref{sec:evaluation}---the number of matched bytes per skip-and-match---impacts the performance, we slightly modify \sysname implementation to support up to 32-byte matching per skip-and-match, by reducing the number of entries in RAM and CAM to run at 250\si{MHz}, which only impacts the configured number of dispatch rules but not impact \sysname hardware performance. Here, we use 256-byte messages in the PingPong application (\S\ref{ssec:eval_microbench}) for this test.

As shown in Figure~\ref{expfig:qn_hw_sw_chars}, \sysname's performance remains constant. This is because matching on CAM can support a longer sequence as long as it does not exceed CAM width. Also, the performance degradation of software implementation is negligible. The reason may be that matching on 32-byte sequences still does not pollute CPU cache to cause performance degradation.

\end{document}